\documentclass[11pt,a4paper]{article}
\pdfoutput=1
\usepackage{jcappub}

\usepackage{graphicx}
\usepackage{times}%
\usepackage{natbib}
\usepackage{epstopdf}
\usepackage{array}
\usepackage{stfloats}
\usepackage{fixltx2e}
\usepackage{amsmath}
\usepackage{xspace}

\newcommand{\ie}{{i.e.}}
\newcommand{\eg}{{e.g.}}
\newcommand{\gsim}{\,\lower2truept\hbox{${>\atop\hbox{\raise4truept\hbox{$\sim$}}}$}\,}

\def\eg{{\rm e.g.$\,$}}
\def\ie{{\rm i.e.$\,$}}

\newcommand{\be}{\begin{equation}}
\newcommand{\ee}{\end{equation}}
\newcommand{\bea}{\begin{eqnarray}}
\newcommand{\eea}{\end{eqnarray}}

\graphicspath{{./}}

\newcommand{\DM}{Dark Matter\xspace}

\newcommand{\LCDM}{$\Lambda$CDM\xspace}
\newcommand{\abs}[1]{\lvert{#1}\rvert}
\newcommand{\virgolette}[1]{``{#1}''}
\newcommand{\gadget}{\texttt{GADGET-2}\xspace}

\newcommand{\zinco}{\texttt{ZInCo}\xspace}
\newcommand{\mean}[1]{\langle #1 \rangle}

\def\ltsima{$\; \buildrel < \over \sim \;$}
\def\simlt{\lower.5ex\hbox{\ltsima}}
\def\gtsima{$\; \buildrel > \over \sim \;$}
\def\simgt{\lower.5ex\hbox{\gtsima}}

\title{Zoomed high-resolution simulations of Multi-coupled Dark Energy: cored galaxy density profiles at high redshift}

\author[1,2]{Enrico Garaldi, }
\author[2,3,4]{Marco Baldi, }
\author[2,3,4]{Lauro Moscardini}

\affiliation[1]{Argelander Institut f\"ur Astronomie der Universit\"at Bonn, Auf dem H\"ugel 71, Bonn, D-53121, Germany}
\affiliation[2]{Dipartimento di Fisica e Astronomia, Alma Mater Studiorum Universit\`a di Bologna, viale Berti Pichat, 6/2, I-40127 Bologna, Italy}
\affiliation[3]{INAF - Osservatorio Astronomico di Bologna, via Ranzani 1, I-40127 Bologna, Italy}
\affiliation[4]{INFN - Sezione di Bologna, viale Berti Pichat 6/2, I-40127 Bologna, Italy}

\emailAdd{egaraldi@uni-bonn.de}
\emailAdd{marco.baldi5@unibo.it}
\emailAdd{lauro.moscardini@unibo.it}

\abstract{
We perform for the first time high-resolution \emph{zoom-in} re-simulations of individual halos in the context of the Multi-coupled Dark Energy (McDE) scenario, which is characterised by the existence of two distinct Dark Matter particle species with opposite couplings to a Dark Energy scalar field. We compare the structural properties of the simulated halos to the standard $\Lambda $CDM results. The \emph{zoomed-in} initial conditions are set up using a specifically designed code called \zinco that we publicly release along with the present paper. Our numerical results allow to investigate in detail and with unprecedented resolution the {\em halo segregation} process that characterises McDE cosmologies from its very early stages. In particular, we find that in contrast to what could be inferred from previous numerical analysis at lower resolution, the segregation process is already in place at redshifts as high as $z\sim 7$. Most remarkably, we find that the subsequent evolution of the segregation leads to the formation of cored total matter density profiles with a core size that progressively increases in time. The shape of the cored profiles can be accurately predicted as the superposition of two NFW profiles with an increasing offset, thereby confirming the interpretation of the simulations results in terms of the segregation of the two dark matter components of the halo as a consequence of their different coupling to the Dark Energy field. 
}


\begin{document}
\maketitle

\section{Introduction}
\label{i}

Despite keeping to prove consistent with an ever increasing and ever improving wealth of observational data \citep[][]{Planck_2015_XIII,Planck_016,SDSS-7,Parkinson_etal_2012}, the standard $\Lambda $CDM cosmological model does not
provide a satisfactory explanation of the physics underlying its phenomenological success. The failure in detecting any of the cold dark matter (CDM) particle candidates through both direct and indirect detection experiments \citep[see e.g.][and references therein]{Bergstrom_2012} after more than two decades of attempts makes the foundations of the CDM paradigm less secure. On the other hand, the extremely low energy scale associated with the cosmological constant $\Lambda $ poses severe naturalness issues, generally known as the {\em fine-tuning} and the {\em coincidence} problems \citep[see e.g.][]{Weinberg_1989,Weinberg_2000}, that can be addressed only by appealing to anthropic arguments \citep[see e.g.][]{Barrow_Tipler_1988}. Moreover, the \emph{satellite abundance} \citep[][]{Moore_etal_1999,Klypin_etal_1999}, the \emph{cusp-core} \citep[][]{deNaray_McGaugh_deBlok_2008,deBlok_2010,Gentile_etal_2004}{, and the \emph{``too big to fail"} \citep[][]{BoylanKolchin_Bullock_Kaplinghat_2011,BoylanKolchin_Bullock_Kaplinghat_2012}} problems have been frequently interpreted as failures of the $\Lambda $CDM model on small scales.

Alternatives to the standard model have thus been investigated with respect to both these two fundamental ingredients by exploring e.g. Warm Dark Matter \citep[WDM, ][]{Bode_Ostriker_Turok_2001}, axions \citep[][]{Peccei_Quinn_1977}, 
non-thermal relics \citep[like sterile neutrinos, see][]{Dodelson_Widrow_1994}, wavelike \citep[]{Medvedev_2014} and flavour-mixed \citep[]{Schive_Chiueh_Broadhurst_2014} Dark Matter, self-interacting Dark Matter \citep[]{Spergel_Steinhardt_2000,Dave_Spergel_Steinhardt_Wandelt_2001,Vogelsberger_Zavala_Simpson_Jenkins_2014,Fan_Katz_Randall_Reece_2013}   or composite Dark Matter \citep[see \eg ][]{Khlopov_2015} 
 as a substitute of CDM, while dynamical Dark Energy \citep[][]{Wetterich_1988,Ratra_Peebles_1988}, interacting Dark Energy \citep[][]{Wetterich_1995,Amendola_2000,Amendola_2004,Farrar2004,Amendola_Baldi_Wetterich_2008,Baldi_2011a}, or low-energy modifications of General Relativity \citep[see e.g.][]{Hu_Sawicki_2007} have been considered as a replacement for the cosmological constant. 

As most of these alternative scenarios, especially when modelling the Dark Energy component, introduce additional degrees of freedom and additional parameters, one of the most widely invoked arguments in favour of the cosmological constant is given by its extreme simplicity as being fully characterised by a single parameter. Nonetheless, some of the above-mentioned (and more physically motivated) models require only a few extra parameters as compared to $\Lambda $CDM and could then stand as reasonable competitors to the standard cosmological paradigm. 

The continuous improvement of the quality of observational data has allowed to rule out entire classes of alternative cosmological scenarios, and to tightly constrain the allowed parameter space of the relatively few surviving models. In this respect, the upcoming decade -- characterised by percent-precision measurements of several cosmological observables -- will provide the opportunity to further improve these constraints to the level where potentially observable features would be ever hardly detectable. Therefore, such quantum leap in the quality of observational data will either detect some deviation from the expectations of $\Lambda $CDM or restrict the allowed models to be basically indistinguishable from the standard scenario over the range of scales and epochs that will be tested by the next generation of large surveys.

In this respect, it is particularly interesting to investigate possible alternative models expected to appear indistinguishable from $\Lambda $CDM in the appropriate regimes that are (or soon will be) tightly constrained by observations, that nonetheless might still allow for potentially detectable deviations elsewhere. In particular, a significant attention has been devoted in recent years to models characterised by various types of {\em screening mechanisms}, that allow to recover the $\Lambda $CDM behaviour under some specific conditions. 

In the context of Modified Gravity theories, for instance, 
the screening allows to recover standard gravity in overdense regions of the universe through a variety of mechanisms \citep[as the {\em chameleon}, {\em symmetron}, and {\em Vainshtein} mechanisms, see e.g.][respectively]{Khoury_Weltman_2004,Hinterbichler_Khoury_2010,Deffayet_etal_2002}, allowing consistency with both cosmological observations and Solar System tests of gravity \citep[][]{Bertotti_Iess_Tortora_2003, Will_2005, Amendola_Tsujikawa_2008, Capozziello_Tsujikawa_2008}.

Alternatively, inspired by Quantum Theory, a number of different models have been proposed, such as the Multi-component Dark Matter in which the Dark Matter particles are quantum flavour-mixed superpositions of mass eigenstates \citep[][studied by means of numerical N-body simulations]{Medvedev_2014}, or a wavelike Dark Matter interpretation \citep[]{Schive_Chiueh_Broadhurst_2014}, in which the bosonic Dark Matter particles form Bose-Einstein condensates which modify their small-scale properties, leaving the large-scale behaviour unchanged with respect to the \LCDM model.

The approach is entirely different for what concerns the vast range of interacting Dark Energy cosmologies \citep[see e.g.][for a general classification]{Pourtsidou_Skordis_Copeland_2013,Skordis_Pourtsidou_Copeland_2015}, including e.g. the widely studied {\em Coupled Quintessence} scenario \citep[see e.g.][]{Amendola_2000,Amendola_2004}, the {\em Growing Neutrino} model \citep[][]{Amendola_Baldi_Wetterich_2008}, and the {\em Dark Scattering} model \citep[][]{Simpson_2010,Baldi_Simpson_2015}. In this class of scenarios, characterised by a direct interaction between a Dark Energy scalar field and massive particles, the local tests of gravity are evaded by allowing the interaction only for \DM  (and/or neutrinos), thereby leaving the baryons completely uncoupled \citep[][]{Damour_Gibbons_Gundlach_1990}. However, the interaction can still have an impact on the expansion history of the universe and on the evolution {of structure formation, in particular by leaving characteristic footprints on e.g. matter clustering statistics \citep[][]{Marulli_Baldi_Moscardini_2012,Moresco_etal_2014,Carlesi_etal_2014a,Casas_etal_2015}, cluster counts \citep[][]{Baldi_Pettorino_2011,Baldi_2011c,Cui_Baldi_Borgani_2012}, large-scale velocity fields \citep[][]{Lee_Baldi_2011}, gravitational lensing \citep[][]{Beynon_etal_2012,Carbone_etal_2013,Giocoli_etal_2015,Pace_etal_2015}, voids statistics \citep[][]{Li_2011,Pollina_etal_2015,Sutter_etal_2015}, and structural properties of collapsed halos \citep[][]{Baldi_Salucci_2011,Baldi_Lee_Maccio_2011,Giocoli_etal_2012,Carlesi_etal_2014b,Penzo_etal_2015}. The detailed predictions of such effects (also by means of numerical simulations) have progressively allowed to use observations for constraining the coupling strength \citep[][]{Amendola_etal_2003,DiPorto_Amendola_2008,Xia_2009,Pettorino_etal_2012,Pettorino_2013,Xia_2013} to a level that makes possible observational signatures at small scales hardly detectable.}\\

Some simple extensions of the original interacting Dark Energy framework have been then proposed with the goal of building plausible alternatives to the cosmological constant featuring testable predictions at the level of  linear and non-linear structure formation without substantially affecting cosmology at larger scales. One of such possible extensions consists in considering a time evolution of the interaction strength \citep[][]{Amendola_2004,Baldi_2011a}.
Alternatively self-interacting Dark Matter \citep[]{Spergel_Steinhardt_2000} has been proposed as a solution for the small-scale issues of the \LCDM cosmology and it was subsequently shown \citep[]{Dave_Spergel_Steinhardt_Wandelt_2001,Vogelsberger_Zavala_Simpson_Jenkins_2014} that the Dark Matter self-interaction affects also the baryonic component, inducing differences with respect to the \LCDM cosmology as large as 50\%. In the Double-Disk Dark Matter \citep[]{Fan_Katz_Randall_Reece_2013} model a two-component Dark Matter, where just the most abundant one is collisionless, is found to be able to explain the satellite plane found in the Milky Way and Andromeda galaxies \citep[]{Metz_Kroupa_Jerjen_2009} and leave space for testable predictions, especially if the Dark Matter annihilates into gamma rays.
A completely different approach, that will be the focus of the present work, amounts to considering multiple species of CDM particles interacting with the Dark Energy field through individual coupling functions. Indeed, a multi-particle nature of the CDM component has been investigated in various theoretical contexts \citep[see e.g. the recent work by][]{Chialva_Dev_Mazumdar_2013}. For instance, a multi-particle dark matter model characterised by a cold and a hot component has been explored \citep[see e.g.][]{Anderhalden_etal_2012,Maccio_etal_2013} as a possible solution to the small-scale issues of the $\Lambda $CDM scenario, considering in some cases also a coupling to a Dark Energy field \citep[][]{Bonometto_Sassi_LaVacca_2012,Bonometto_Mainini_2014,Bonometto_etal_2015,Maccio_etal_2015}.
Furthermore, theoretical models featuring multiple CDM species with different couplings to a Dark Energy scalar have been proposed by \cite{Gubser_Peebles_2004} and subsequently extensively discussed by
\cite{Gubser2004,Farrar2004,Nusser_Gubser_Peebles_2005}. 

The simplest realisation of this general class of models has been recently investigated by \cite{Baldi_2012a} and termed ``The Multi-Coupled Dark Energy" scenario (McDE, hereafter). The latter is characterised by only two distinct CDM particle species with opposite constant couplings to a Dark Energy scalar field, thereby requiring no additional free parameters as compared to standard Coupled Quintessence, and features attractor solutions in the background evolution \citep[][]{Brookfield_vandeBruck_Hall_2008} that suppress the effective coupling during matter domination, providing a sort of {\em background screening mechanism} \citep[][]{Baldi_2012a,Piloyan_etal_2013}. As a consequence of the opposite couplings of the two CDM components, the McDE scenario will feature both attractive and {\em repulsive} fifth-forces between CDM particles, differently from all standard interacting Dark Energy models where fifth-forces are always attractive. This represents the most distinctive feature of the McDE model, and gives rise to a highly non-trivial phenomenology at the level of linear and non-linear structure formation. 

Such phenomenology has already been investigated both at the level of linear perturbations \citep[][]{Piloyan_etal_2014} -- showing that linear growth suddenly deviates from the $\Lambda $CDM behaviour when the coupling exceeds the threshold corresponding to a fifth-force as strong as standard gravity -- and at the level of non-linear evolution through a set of suitably designed N-body simulations {by e.g. \citep[][B13 hereafter]{Baldi_2013} and \citep[][]{Baldi_2014}}. The latter have highlighted for the first time the phenomenon of {\em halo segregation}, occurring in McDE cosmologies as a consequence of the repulsive interaction between CDM particles of opposite types and of their different mass evolution resulting in an effective violation of the Weak Equivalence Principle.

In the present work, we aim to investigate in detail the {\em halo segregation} process by focusing on individual CDM halos through high-resolution {\em zoom-in} simulations following the segregation from its early stages until the original structure (initially composed of a mixture of the two CDM species) finally splits into two separate halos dominated by a single CDM type. In particular, we study how the total CDM density profile, 
which can be observed 
\eg through gravitational lensing or tested by the motion of baryonic test particles (like stars), evolves during the segregation process. {For setting up the initial conditions of our simulations we make use of a new code called \zinco (Zoomed INitial COnditions), that we publicly release along with the present work.}\\

The paper is organised as follows.
In Section \ref{models} we briefly review the main equations of the McDE scenario. In Section \ref{zsims} we describe the numerical setup for our {\em zoom-in} simulations. In Section \ref{results} we present our results, with a particular focus on the halo density profiles during the segregation process. In Section \ref{amodel} we present an analytical model describing the total profile during the segregation. Finally, in Section \ref{concl} we draw our conclusions{, while in the Appendix \ref{ZInCo} we provide a detailed description of the \zinco algorithm and we outline the procedures to access and use the code}.

\section{Multi-coupled Dark Energy models}
\label{models}

Here we provide a brief description of the McDE scenario, by describing the main equations that govern the model and by introducing the notation that will be used throughout the paper. We refer to \citep[][]{Baldi_2012a,Piloyan_etal_2013} for a more thorough discussion on the model.\\

The background evolution of McDE cosmologies is described by the following set of equations:
\begin{eqnarray}
\label{klein_gordon}
\ddot{\phi } + 3H\dot{\phi } + \frac{dV}{d\phi } &=& +C \rho _{+} - C \rho _{-}\,, \\
\label{continuity_plus}
\dot{\rho }_{+} + 3H\rho _{+} &=& -C \dot{\phi }\rho _{+} \,, \\
\label{continuity_minus}
\dot{\rho }_{-} + 3H\rho _{-} &=& +C \dot{\phi }\rho _{-} \,, \\
\label{continuity_radiation}
\dot{\rho }_{r} + 4H\rho _{r} &=& 0\,, \\
\label{friedmann}
3H^{2} &=& \frac{1}{M_{{\rm Pl}}^{2}}\left( \rho _{r} + \rho _{+} + \rho _{-} + \rho _{\phi} \right)\,,
\end{eqnarray}
where the scalar field $\phi $ plays the role of Dark Energy, the $\pm $ subscripts denote
the two different CDM species according to the sign of the coupling to the DE scalar field, the subscript $r$ indicates relativistic particles, and C is the coupling strength, 
usually
expressed in its dimensionless form $\beta $:
\begin{equation}
\beta \equiv \sqrt{\frac{3}{2}}M_{\rm Pl}C \,,
\end{equation}
with $M_{\rm Pl} \equiv 1/\sqrt{8\pi G}$ being the reduced Planck mass and  $G$ the Newton's constant.
In Eqs.~\ref{klein_gordon}-\ref{friedmann} an overdot indicates a derivative with respect to the cosmic time $t$, $H$ is the Hubble function, and the scalar self-interaction potential is assumed to take the exponential form \citep[][]{Lucchin_Matarrese_1984,Wetterich_1988,Ferreira_Joyce_1998}:
 \begin{equation}
V(\phi ) = Ae^{-\alpha \phi /M_{\rm Pl}}\,. 
\end{equation}

Any dynamical evolution of the Dark Energy scalar field $\dot{\phi} \neq 0$ produces a variation in the mass of \DM particles, according to
\begin{equation}
\label{mass-evol}
\frac{d[m_{\pm} / M_{Pl}]}{dt} = \mp C \dot{\phi}\ ,
\end{equation}
where $m_{\pm}$ is the mass of a positively- or negatively-coupled \DM particle.

The above system of equations is characterised by an attractor solution during matter domination where the two  CDM particle species evenly share the total energy budget of the universe \citep[see][]{Piloyan_etal_2013}. In such regime, Eq.~\ref{klein_gordon} clearly shows that the effective coupling is strongly 
suppressed and the cosmic background evolution is indistinguishable from that of an uncoupled system. As the attractor is unstable, the system will eventually evolve to a Dark Energy dominated solution, developing an asymmetry between the two CDM species and giving rise
to a non-vanishing effective coupling at late times.\\

At the level of sub-horizon linear density perturbations, the opposite coupling constants of the two different CDM particle types introduce a few correction terms to the standard gravitational instability equations.
The evolution equations for the linear density contrast at sub-horizon scales in McDE take the following form:
\begin{eqnarray}
\label{gf_plus}
\ddot{\delta }_{+} = -2H\left[ 1 - \beta \frac{\dot{\phi }}{H\sqrt{6}}\right] \dot{\delta }_{+} + 4\pi G \left[ \rho _{-}\delta _{-} \Gamma_{R} + \rho _{+}\delta _{+}\Gamma_{A}\right] \,, \\
\label{gf_minus}
\ddot{\delta }_{-} = -2H\left[ 1 + \beta \frac{\dot{\phi }}{H\sqrt{6}}\right] \dot{\delta }_{-} + 4\pi G \left[ \rho _{-}\delta _{-} \Gamma _{A} + \rho _{+}\delta _{+}\Gamma_{R}\right]\,,
\end{eqnarray}
where: 
\begin{equation}
\label{def_gamma}
\Gamma _{A} \equiv 1 + \frac{4}{3}\beta ^{2}\,, \quad \Gamma _{R}\equiv 1 - \frac{4}{3}\beta ^{2} \,,
\end{equation}
represent attractive ($\Gamma _{A}$) or repulsive ($\Gamma _{R}$) corrections to gravity due to the long-range fifth-force.
Again, we can notice that for $\rho _{+} = \rho _{-}$, which is realised during the matter-dominated attractor, and assuming adiabatic initial conditions (i.e. $\rho _{+}\delta _{+} = \rho _{-}\delta _{-}$) the $\Gamma _{\pm }$ corrections exactly vanish in both equations, so that the standard gravitational source term is effectively restored.
However, the second terms in the first square brackets on the right-hand side of Eqs.~\ref{gf_plus}-\ref{gf_minus}, representing two opposite drag terms for the density perturbations of the two different CDM species, will tend to move the system away from adiabaticity whenever the Dark Energy field evolves in time (\textbf{\ie} $\dot{\phi } \ne 0$), thereby triggering the fifth-force corrections also along the attractor, since even if $\rho _{+}=\rho _{-}$,  the $\Gamma $ factors are non-vanishing for $\delta _{+}\neq \delta _{-}$. Nonetheless, it was shown by \citep[][]{Baldi_2012a,Piloyan_etal_2013} that the deviation from adiabaticity always remains small along the attractor for coupling values $\beta \leq \beta _{G} \equiv \sqrt{3}/2$, corresponding to the value at which the fifth-forces become comparable to standard gravity. Such {\em ``gravitational threshold"} represents a clear barrier for McDE models: for lower couplings the resulting cosmological evolution is very similar to the $\Lambda $CDM case even at the level of non-linear structure formation, while for larger coupling values linear perturbations show a steeply enhanced growth \citep[][]{Piloyan_etal_2014} while non-linear structures are drifted apart by the repulsive fifth force \citep[][]{Baldi_2013,Baldi_2014} giving rise to the halo segregation effect. In fact, the threshold between these two regimes $\beta = \beta _{G}$ represents the only case for which the repulsive fifth-force acting between two CDM particles of opposite type exactly cancels their mutual gravitational attraction, so that particles of opposite type do not exert any force on each other. This special condition will be the focus of our numerical investigation of non-linear structure formation in McDE.

\subsection{Results from previous simulations}
\label{psims}
Non-linear structure formation in the McDE model has been studied {in B13 and \citep[][]{Baldi_2014}} using cosmological N-body simulations with homogeneous resolution over the simulation box. These works confirmed the qualitative behaviour (already tested at the level of linear perturbations evolution) that for coupling strengths below $\beta _{G}$ the formation of cosmic structures is very mildly affected by the interaction. Additionally, these first numerical investigations also showed that higher values of the coupling result in the formation of \emph{mirror} structures, \ie pairs of halos clearly separated from each other, with comparable masses, and dominated by opposite types of \DM particles; the spatial offset between the two objects was shown to grow with time in the direction of their peculiar velocity. 
Therefore, as such mirror structures are not observed in real data, these numerical results restrict the viable range of the coupling $\beta$ to $\beta \lesssim \beta _{G}$. On the other hand, since the most peculiar and most relevant features of the McDE cosmology start to become appreciable only for coupling strengths near such threshold, the range of parameters that make the model both phenomenologically interesting and observationally viable is restricted to $\beta _{G}-\epsilon < \beta \leq \beta _{G}$, with $\epsilon $ to be determined empirically. 

In order to fully understand the segregation mechanism we therefore choose to focus the present analysis on the special case $\beta = \beta _{G}$, which represents a unique and particularly simplified situation as particles of different \DM species feel a vanishing net force (\ie they do not directly interact with each other in any way), as already explained above. As a consequence, all the features that will appear in our simulations will have to be addressed to the only surviving effect in the perturbed dynamical equations \ref{gf_plus}-\ref{gf_minus}, \ie the different friction terms of the two CDM species, thereby greatly simplifying the interpretation of the results. A deeper understanding of the segregation process for such simplified setup will certainly help in future analysis of models featuring a lower coupling strength.
In all our models the slope $\alpha $ {of the} self-interaction potential was set to $\alpha = 0.08$, {which allows to keep the fraction of Early Dark Energy ($\Omega _{\phi ,{\rm e}}\approx n/\alpha ^{2}$, with $n=3$ for matter domination and $n=4$ for radiation domination) within presently allowed bounds \citep[][]{Planck_2015_XIV}}. Previous works have shown that the effects of McDE on structure formation are in any case {largely} insensitive to the exact value of the potential slope $\alpha$ \cite{Baldi_2012b}.

\section{Zoomed N-body simulations}
\label{zsims}

In order to investigate in detail the process of {\em halo segregation} during its developments for a McDE model with a coupling at the gravitational threshold $\beta _{G}$ we have extended previous works by running for the first time zoomed high-resolution re-simulations of individual collapsed halos undergoing the segregation process. We now describe in more detail the simulations performed for this work and their analysis pipeline.

\subsection{Simulations details}
\label{sim_details}

\begin{table}[t]
\centering
\begin{tabular}{l|cccccc}
name & \parbox{2.2cm}{\centering $m_{+}^{hr}(z_i)$ \\ $(10^7 M_{\odot})$} & \parbox{2.2cm}{\centering $m_{-}^{hr}(z_i)$ \\ $(10^7 M_{\odot})$} & $z_i$ & $z_f$ & \parbox{1.5cm}{\centering $\epsilon_g^{hr}$ \\ $(kpc/h)$} & $z$-focus \\
\hline
\emph{LR-Dilution} & $\approx 117.7$ & $\approx 111.9$ & $99$ & $0$ & $8$ & $1.2-0$ \\
\emph{LR-Zoom} & $1.8$ & $1.7$ & $99$ & $0$ & $2$ & $1.2-0$ \\
\emph{HR-Dilution} & $\approx 117.7$ & $\approx 111.9$ & $99$ & $4$ & $8$ & $7-4$ \\
\emph{HR-Zoom1} & $0.226$ & $0.222$ & $99$ & $4$ & $1$ & $7-4$ \\
\emph{HR-Zoom2} & $0.226$ & $0.222$ & $99$ & $4$ & $1$ & $7-4$ \\
\end{tabular}
\caption{Main properties of the simulations presented in this paper. Each item represents two different simulations with the same parameters but different cosmological models: McDE and \LCDM. $m_{\pm}^{hr}(z_i)$ is the \emph{initial} mass resolution for the high-resolution region in the simulation for the positively- and -negatively coupled particles respectively; $z_i$ and $z_f$ are the initial and final redshifts of the simulation, respectively; $\epsilon^{hr}$ is the minimum value of the softening length used, \ie the value used for the high-resolution particles and $z$-focus is the redshift range where the code was set up to produce a high density of snapshots. The \virgolette{$\approx$} symbol emphasises that $m_{\pm}^{hr}(z_i)$ is a \emph{mean} mass resolution due to the algorithm used to produce the ICs, which does not ensure the number of merged particles to be exactly constant. The prefixes \emph{LR} and \emph{HR} denote the lower-resolution (\ie $2 \times 1024^3$ particles for each \DM species) and higher-resolution (\ie $2 \times 2048^3$ particles for each \DM species) run, respectively. \emph{Dilution} refers to a uniform downsampling of the original high resolution ICs, used as a preparatory run, while \emph{Zoom} refers to the zoom-in simulations on a specific region of interest.}
\label{table_sim}
\end{table}

We carried out two different sets of simulations using the cosmological N-body code {\small C-GADGET} \citep[][]{Baldi_etal_2010}, a modified version of the widely-used TreePM N-body code {\small GADGET2} \citep[][]{gadget-2}. More specifically, a first set of low-resolution simulations has been used as a testbed for the physical processes to be tracked by a subsequent set of higher resolution simulations that have been used to perform the actual analysis. The main features of these various simulations are summarised in Table \ref{table_sim}, where they are labelled using prefixes \emph{LR} and \emph{HR} for the former and the latter, respectively. 
The \emph{LR} simulations were calibrated following the results of previous numerical studies, summarised in Section \ref{psims}. More specifically, they were set up in order to study with a high time resolution the redshift range between $z=2$ and $z=0.5$, where the segregation process was expected to occur based on the outcomes of \cite{Baldi_2014}. However, our \emph{LR} simulations unexpectedly showed for the first time that the segregation actually starts at much earlier epochs than it was previously inferred due to the insufficient resolution employed in previous studies. Consequently, the \emph{HR} set of simulations was set up based on the new finding of an early onset of the segregation process, thereby focusing a high temporal resolution (i.e. a frequent dump of simulation snapshots) in the higher redshift range $7\ge z \ge 4$. 
Details on such early segregation are described in Section \ref{results} using the \emph{HR} set. Since the \emph{LR} set of simulations was used only to tweak the \emph{HR} simulations, we will refer only to the latter in the rest of the paper, where not differently stated.

The \emph{HR} simulations feature a box side of $80$ Mpc $/ h$ and were evolved starting from redshift $z_i = 99$ down to redshift $z_f = 4$. A preparatory low-resolution version of these simulations with homogeneous resolution (named \emph{HR-Dilution} in Table \ref{table_sim}) was run in order to identify structures suitable to be studied by means of a subsequent zoom-in run. The high-resolution zoom-in simulations (labelled \emph{HR-Zoom1} and \emph{HR-Zoom2} in Table \ref{table_sim}) feature three different resolution regions, with highest initial\footnote{Note that the mass resolution changes with the redshift since the \DM particle mass evolves as described by Eq. \ref{mass-evol}} mass resolutions equal to $m_{+} = 2.26 \cdot 10^6 M_{\odot} / h$ and $m_{-} = 2.22 \cdot 10^6 M_{\odot} / h $ for positively- and negatively-coupled species, respectively.

The initial conditions for these simulations were produced using the code \zinco, developed to manipulate existing initial conditions and described in Appendix \ref{ZInCo}. A \emph{single} set of high-resolution ICs featuring the WMAP7 cosmological parameters \cite{wmap7} was processed in order to produce different resolution configurations of the same initial density field. The gravitational softening $\epsilon_g$ was set to approximatively $1/40$-th of the mean inter-particle distance for all the regions, reaching a minimum value of $1 kpc/h$ for the highest-resolution particles. Each of these simulations was performed both for the standard \LCDM cosmology (the \emph{uncoupled case}\footnote{In our work, the \LCDM was implemented as a McDE with vanishing coupling. Thus, the only difference with a true \LCDM model is given by the Dark Energy, which is mimicked by a scalar field. However, the self-interaction potential for the scalar field was chosen in such a way to be (and proved to be) indistinguishable from the pure \LCDM cosmology \citep[][]{Baldi_2012a}. Moreover our approach allows a fully consistent comparison of the simulation runs, since exactly the same version of the simulation code was used, the only difference being the value of the coupling set.}) and for the McDE cosmology (the \emph{coupled case}) in order to allow a direct comparison of the two models. The simulations were set up in order to produce snapshot every $\Delta z = 0.1$ between $z = 7$ and $z=4$, following the results of the \emph{LR} simulation set.

\subsection{Halos selection and characterization}
\label{halo_selection}

\begin{figure}[!pt]
\begin{minipage}{0.33\textwidth}
\includegraphics[width=\textwidth]{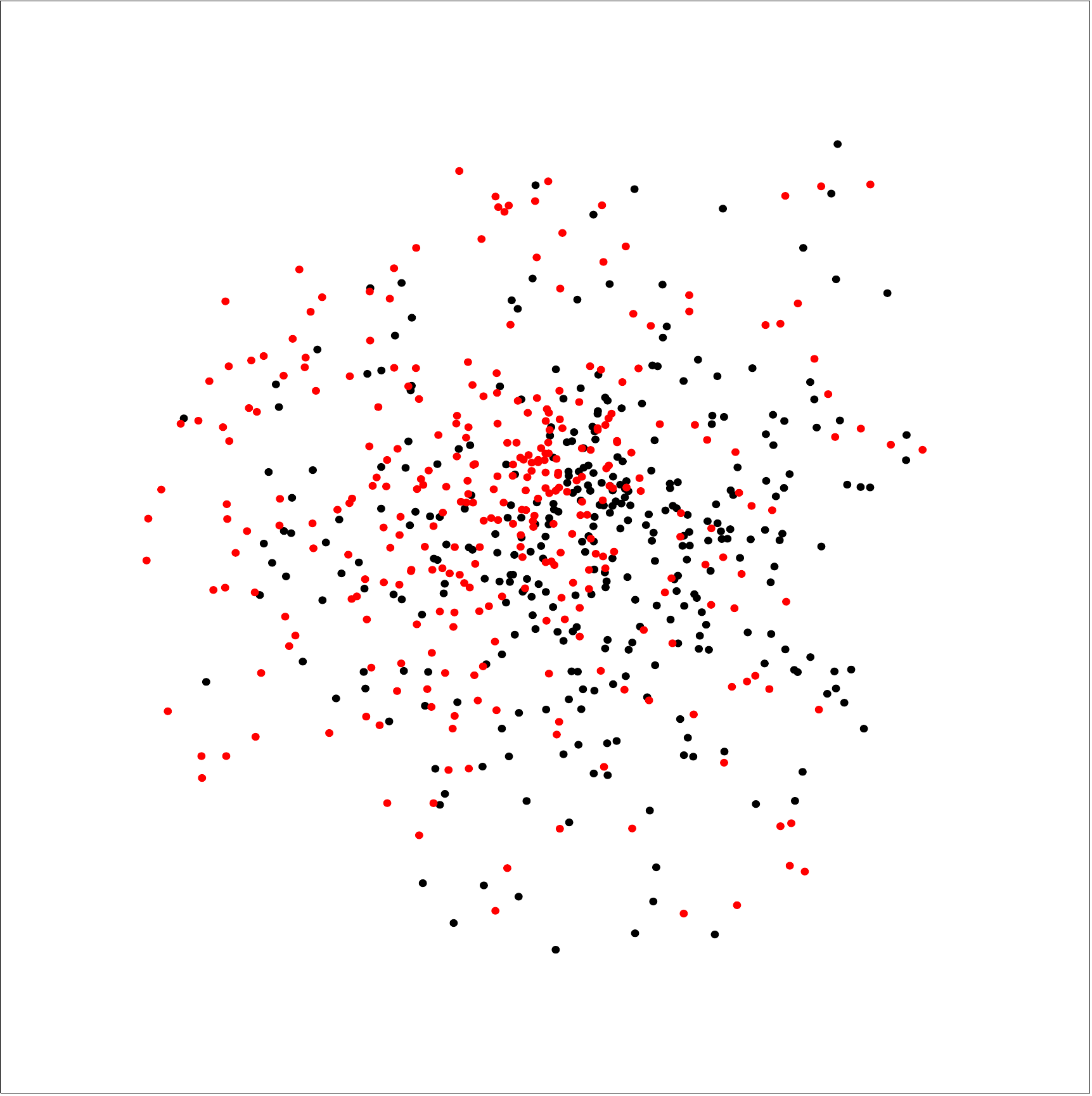}
\end{minipage}
\begin{minipage}{0.33\textwidth}
\includegraphics[width=\textwidth]{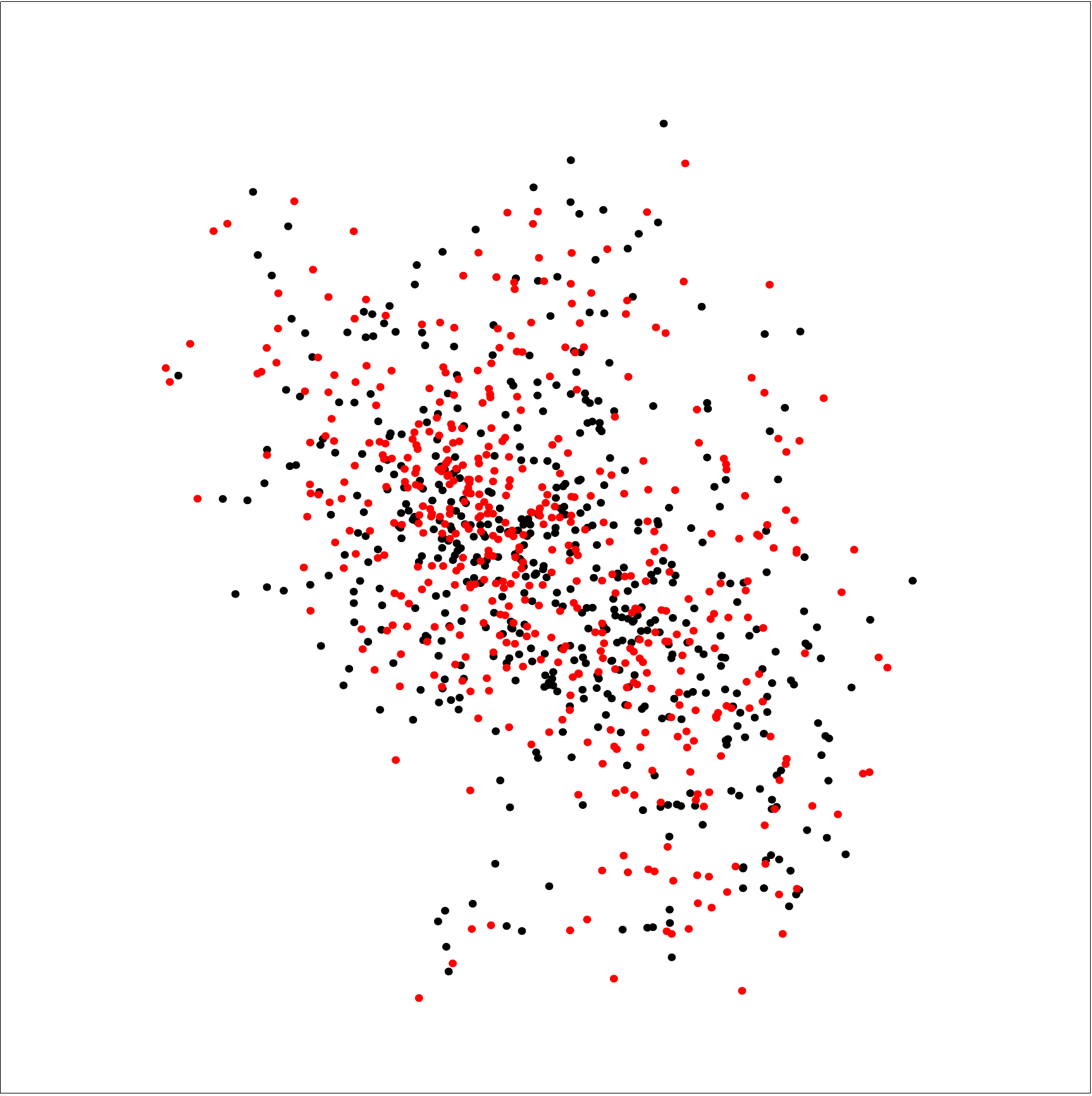}
\end{minipage}
\begin{minipage}{0.33\textwidth}
\includegraphics[width=\textwidth]{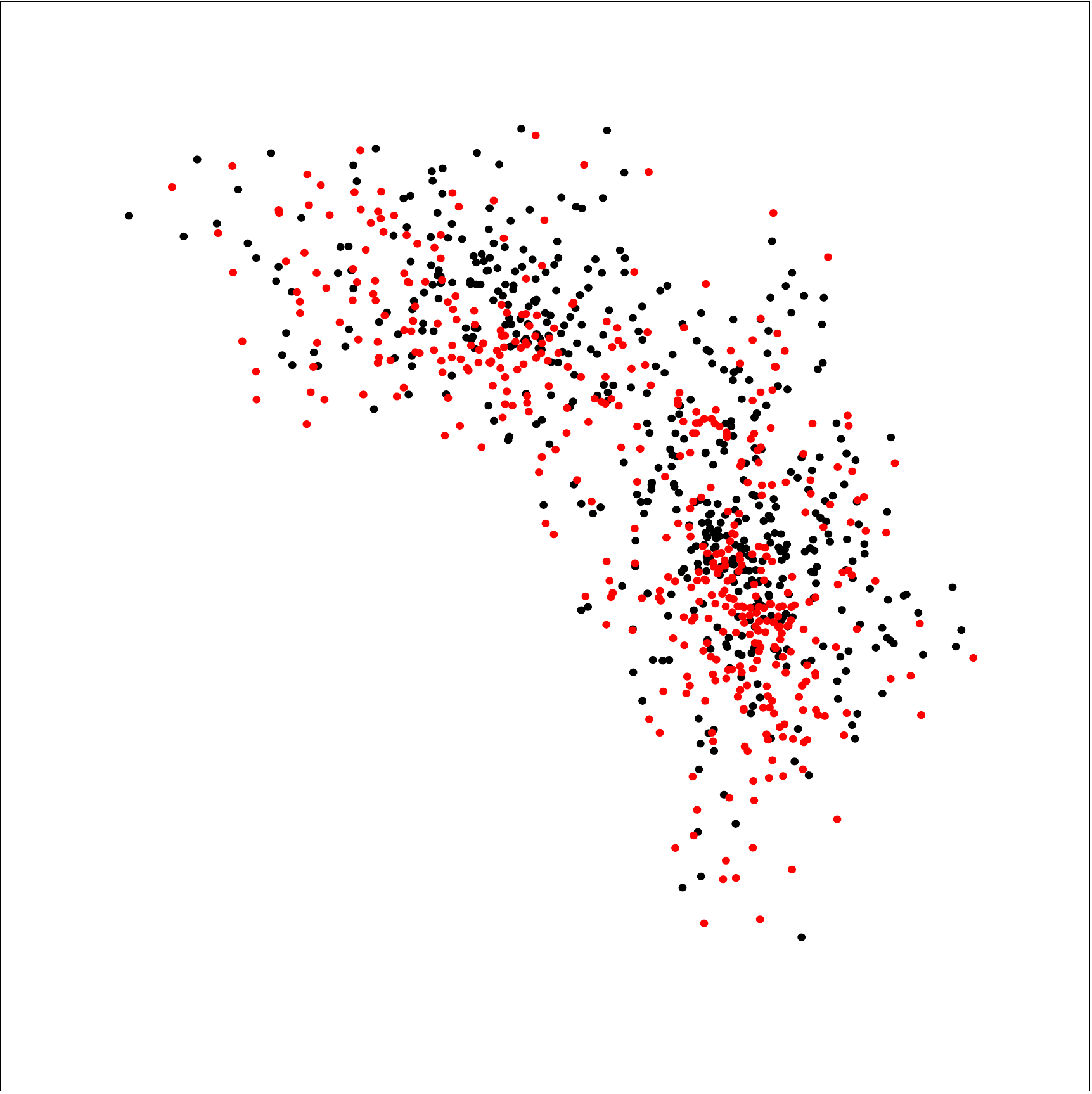}
\end{minipage}
\caption{Examples of selected (left panel) and rejected (middle and right panels) halos. All the plots show the particles distribution of a FoF halo at redshift $z=4$ in the low-resolution (\emph{HR-diluted}) simulation. The two different particle species are plotted using different colors. \emph{Left panel}: a halo showing a visible segregation in two subhalos composed by different \DM species; such segregation can be appreciated as an offset between the two species, with the red-plotted particles on average closer to the left-hand-side of the plot and the black-plotted ones closer to the right-hand-side. \emph{Middle panel}: a halo showing a typical particle distribution but no sign of segregation. \emph{Right panel}: a halo showing a remarkable segregation, but not due to the McDE fifth force since the two substructures are not similar and are not made of different \DM species. }
\label{rejected}
\end{figure}

Once the simulation with reduced resolution (i.e. the \emph{HR-Dilution} run) was performed for both the coupled and the uncoupled models, we performed an initial selection of the halos suitable to be zoomed-in relying on the Friends-of-Friends (FoF) catalogue extracted from the simulation snapshot at $z=4$ using a linking length equal to $0.2$ times the mean inter-particle separation. Then, a further selection was performed by visual inspection of the FoF halos, in order to ensure that the segregation appearing in the McDE cosmology could not be ascribed to effects other than the cosmological model itself. To this end, we compared each suitable structure in the McDE with its counterpart in the \LCDM simulation and we selected only the ones where the substructures are clearly a consequence of the cosmological model, discarding all the halos that at $z=4$ do not show two clear substructures made of different particle species, with similar distribution and remarkable offset. The latter two properties were enforced in order to ensure the segregation is a consequence of the drag term in the McDE model and not addressable to different effects, such as \eg mergers. An example of discarded and selected halos can be found in Figure \ref{rejected}. Given the high number of halos available and the small number of halos to be selected for the zoom-in simulations, such procedure gives reliable results, since the two halos are chosen among the most clearly segregated halos in the simulation. Once a suitable halo was chosen, we followed its evolution from high redshift ($z = 10$) in order to ensure once again that no effect other than the drag term could have produced the observed configuration. From the final pool of suitable halos, we chose the two halos with the most prominent segregation and smooth accretion history and produced zoom-in ICs by means of the \zinco code (see Appendix \ref{ZInCo}). We adopted large high-resolution regions which, besides ensuring no mixing with low-resolution particles, allow to study a number of highly-resolved halos other than the original target of the zoom-in.

When computing the radial density profiles, and in particular when focusing on the features of the inner part of the profile, the choice of the center of the halo with respect to which the profile is computed is crucial, since an offset between the real center and the inferred one could produce a fictional distortion of the profile, possibly turning a cuspy shape into a cored one. In order to determine the center of a \DM halo we used a geometry-based approach, where the center is determined in a recursive way. We chose this method for a consistency reason: since it does not rely on any physical property of the particles, it can be used also in the McDE simulations, whilst the usual group finders would have required to be modified in order to account for the new fifth force typical of the McDE model, producing potentially inconsistent results. Moreover, the algorithm used is almost insensitive to differences in the halo selection, \ie in the exact boundary chosen for the halo.
{The latter is based on a recursive procedure. Given the set of particles belonging to each identified FoF group:} 
\begin{itemize}
\item[{\em i)}] the center of mass (CM) is computed;
\item[{\em ii)}] particles more distant from the CM than a threshold $r_{t}$ are discarded;
\item[{\em iii)}] the new center of mass is computed;
\item[{\em iv)}] $r_{t}$ is decrease by a fixed percentage $f_{limit}$;
\item[{\em v)}] if the number of particles survived is lower than both a threshold number $N_{t}$ and a threshold fraction $f_{t}$ of the original number of particles, the last computed CM is assumed, otherwise the procedure is repeated.
\end{itemize}
We have set $f_{limit} = 0.05$, $N_{t} = 1000$ and $f_{t} = 0.01$ for our analysis.

In the uncoupled case, we apply the algorithm on the whole set of particles selected. However in the coupled case such approach would fail due to the bimodal distribution of the density field in the halo, which leads to the algorithm converging on only one of the two peaks. In order to overcome this problem, we exploit here the fact that the two subhalos are on average composed by different types of \DM particles. Thus, in the coupled case, we apply the algorithm to the two species separately. The halo center is then chosen as the middle point between the two points determined for the two substructures. In the following, when not stated differently, we refer to it as the centre of the halo. Thanks to the fact that the number of particles in a given halo is, on average, equally distributed between the two \DM species, which have very similar mass, the choice of the middle point is almost equivalent to choosing the center of mass of the two subhalos, with differences which are far below the gravitational softening, and thus do not affect the computed profile. Once the center is determined, the profile is computed using logarithmically equispaced radial bins.

\section{Results}
\label{results}
We now present the main results, focusing on the properties of the halos in the McDE cosmology discovered for the first time thanks to the simulations described in Section \ref{zsims}.

As a consequence of conservative choices in the setup of the simulations (in particular a large high-resolution region), we are able to study not only the structure initially targeted for the zoom-in procedure, but also a number of other structures that happen to lie completely in the high-resolution zone, which is approximately a sphere of radius $4~Mpc/h$ in both the high-resolution zoom-in simulations \emph{HR-Zoom1} and \emph{HR-Zoom2}.

The halo selection in the zoom-in simulations was performed following the same procedure described in Section \ref{halo_selection} for the preparatory run. Here we additionally checked that the halos selected are free of contamination from low-resolution particles. Moreover, we selected only halos which contain a sufficiently high number of particles, in order to rule out numerical artifacts: each halo analysed in the zoom-in simulations contains more than $10^4$ particles (and often twice as more) over all the redshift range of interest (\ie from $z=7$ to $z=4$).

\subsection{Early halo segregation}
We start by describing the segregation effect of the \DM halos in the McDE cosmology. As described {in B13} and summarised in Section \ref{models}, the first simulations of the McDE model showed that halos with non-negligible peculiar velocity tend to split at low redshifts into two separate objects dominated by opposite CDM particle species.

\begin{figure}[!pt]
\begin{minipage}{0.33\textwidth}
\centering $z=7$\\
\includegraphics[width=\textwidth]{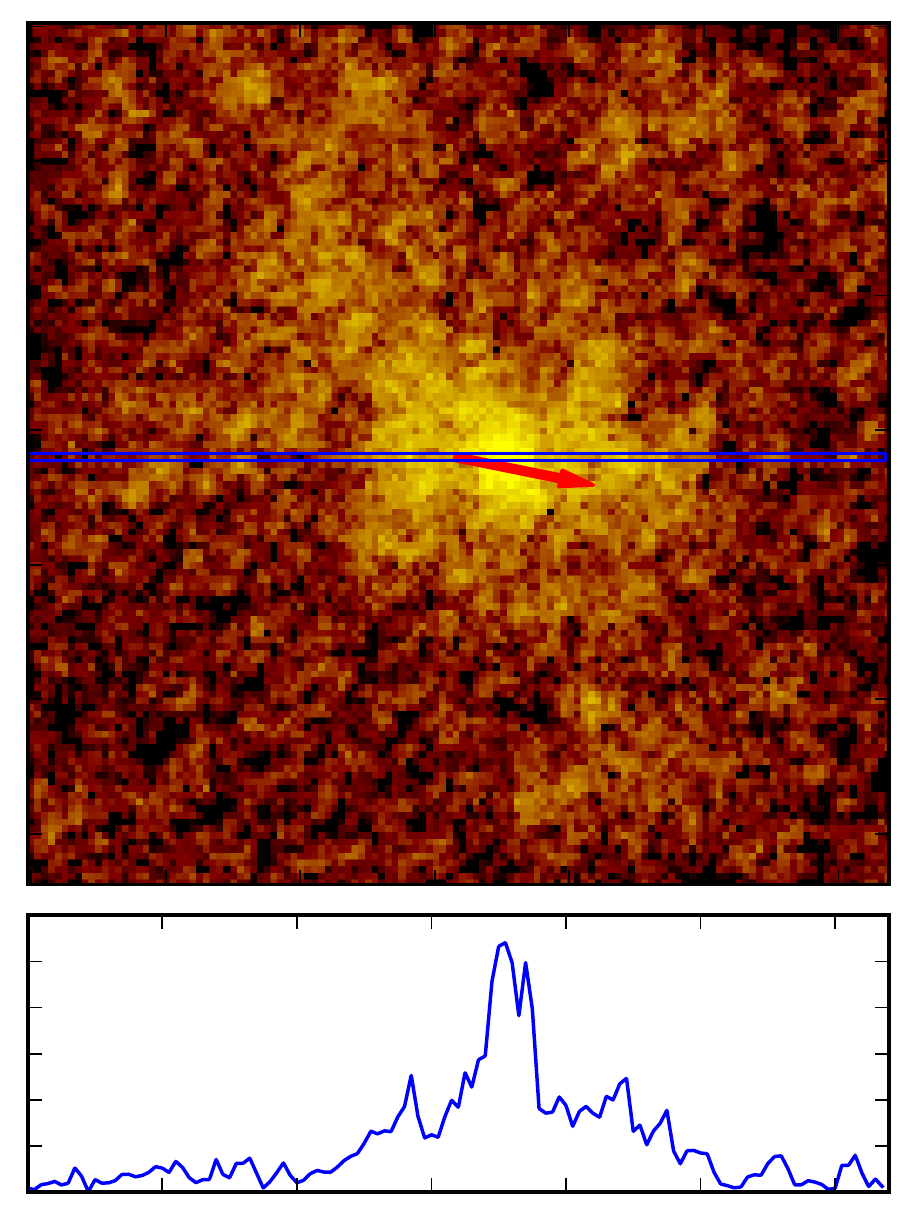}
\end{minipage}
\begin{minipage}{0.33\textwidth}
\centering $z=5$\\
\includegraphics[width=\textwidth]{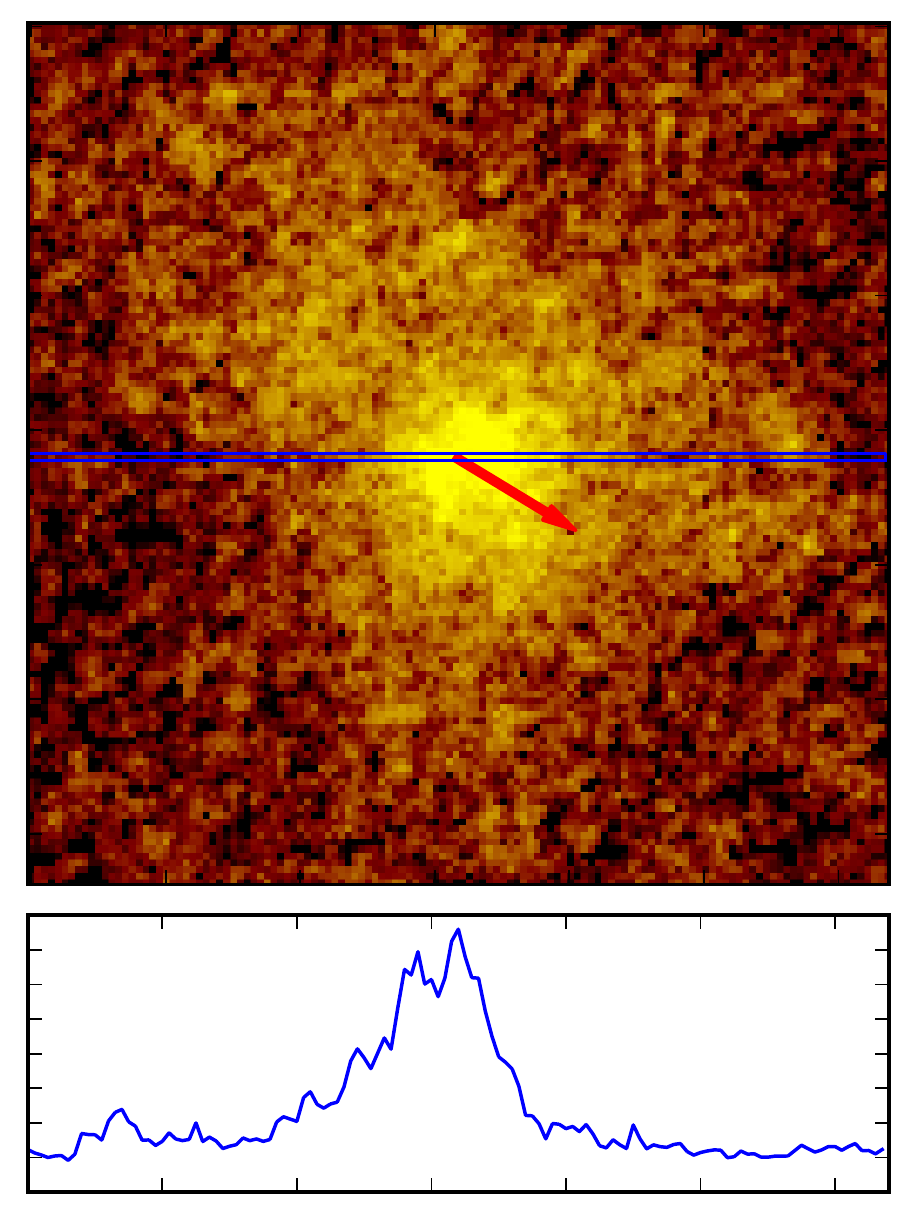}
\end{minipage}
\begin{minipage}{0.33\textwidth}
\centering $z=4$\\
\includegraphics[width=\textwidth]{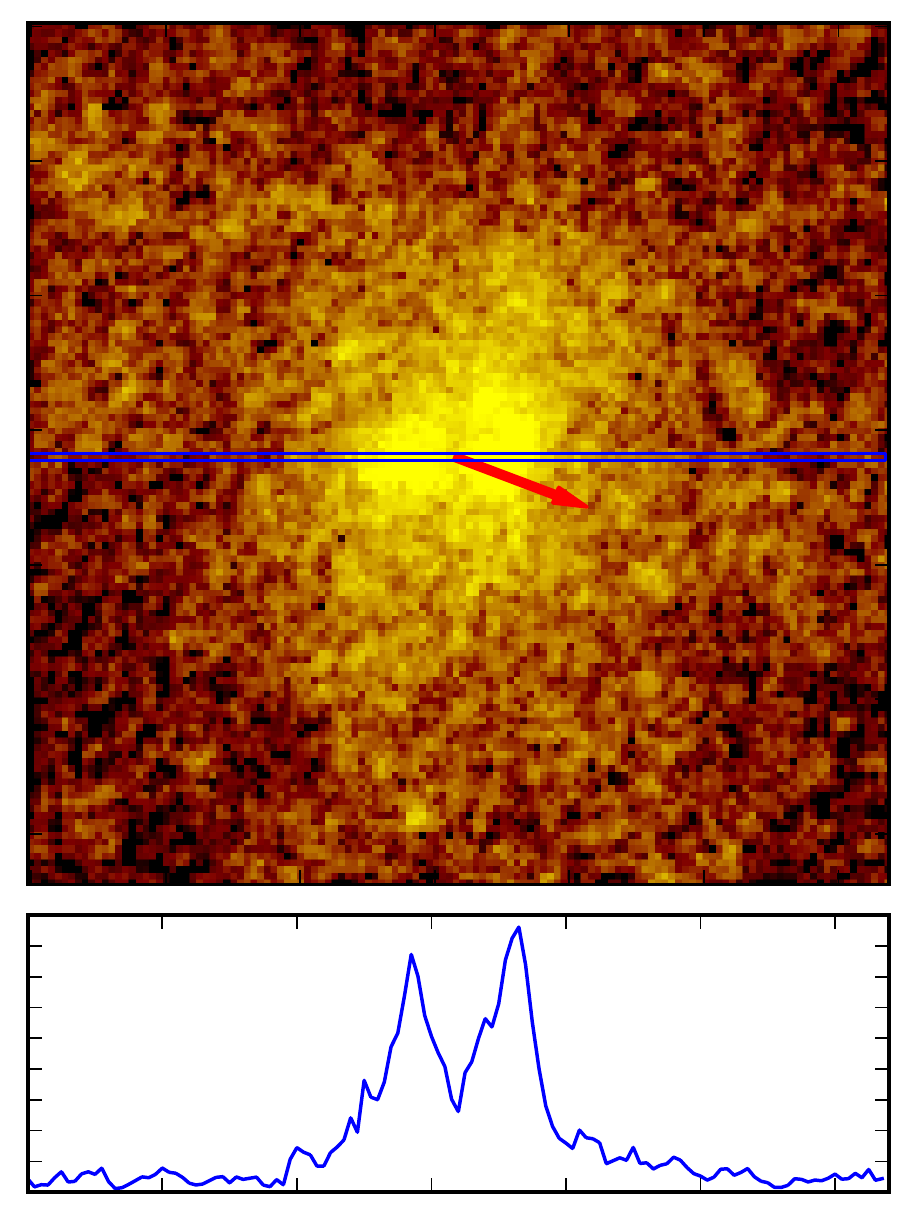}
\end{minipage}
\caption{\emph{Top panels}: Density maps at different redshifts (labels on top of each plot) centred on one of the halos in the high-resolution region of the  McDE simulation. All the plots share the same logarithmic colour scale. \emph{Bottom panels}: Linear density profile (in arbitrary units) along the slits plotted in the above density maps. All the slits pass through the centre of the halo and are aligned along the direction of the centres' offset (rotated to coincide with the $x$ axis). The red arrows represent the direction of the peculiar velocity centre of mass of the halo.}
\label{halo-evolution}
\end{figure}

The evolution of a single halo located in the high-resolution region can be qualitatively appreciated in Figure \ref{halo-evolution}. Note however that all the halos analysed share the same features in the accretion history. In Figure \ref{halo-evolution} a series of density maps (obtained from a slice of the simulation box) at different redshifts are shown. Below each of them a 1-dimensional profile is plotted. The latter is obtained as the pixel intensity (thus proportional to the logarithm of the projected density) along a slit passing through the centre of the halo and properly oriented (shown with blue contour in the top panels). The figures show that even though the two substructures start to appear clearly segregated only at the smallest redshift shown, \ie $z = 4 $, the \virgolette{seed} of such segregation is already present at redshift $z \sim 7$, where the core of the halo shows two close but distinct density peaks. {In B13} the segregation effect was found to become relevant at redshift $z \lesssim 2$, as a consequence of the resolution allowed by that preliminary study, which is low when compared to the one reached by the simulations we performed. Thus, in the simulations described {in B13} the segregation becomes visible only in its advanced phases, when the offset between the two density peaks is sufficiently larger than the numerical spatial resolution. On the contrary, in our simulations the high-resolution allows to distinguish closer density peaks and the segregation appears now to span a wider range of redshifts. 

In fact, all the analysed halos appear to share a similar evolution history: they accrete around two cores, one for each \DM species, with an offset growing with time. Such evolution explains why the previous simulations performed for the McDE cosmological model estimated a much lower segregation redshift: the lower resolution did not allow to resolve the two near-but-separated cores, but they could only catch the later phases of segregation, where the two cores have a bigger offset. Such evolution can be fully understood from the drag terms in Eqs.~\ref{gf_plus}-\ref{gf_minus}, which linearly depend on the density perturbation growth $\dot{\delta }_{\pm}$. The drag determines a small offset along the direction of peculiar motion between the two \DM species already at high redshifts, as soon as the original halo starts to fall in the gravitational potential of the surrounding cosmic structures. The two offset objects then aggregate around different cores. As redshift decreases the drag terms become stronger, producing an increasing offset between the two accretion cores, which are slowly driven apart.

It is important to remark here that -- as a consequence of the particular choice of the McDE coupling $\beta = \beta _{G}$ -- the two halos do not feel each other's presence since their mutual repulsive fifth-force exactly cancels the gravitational attraction, and therefore will grow by accreting CDM particles of a single species as if there were isolated.

It should be noted that we explicitly selected for our analysis halos presenting a clear segregation in the \DM species at redshift $z = 4$, given our main purpose of studying the density profile of \emph{segregated} halos. Thus the halo segregation is certainly addressable as a consequence of the drag force typical of the McDE models, but the accretion history described is not necessarily shared by all the halos in such cosmology, since halos can avoid such segregation \eg if they have very small peculiar velocities or as a consequence of mergers.


\subsection{Mass density profiles}
As a consequence of the segregation process which involves the \DM halos in the McDE cosmology, their total density profile (as could be tested by e.g. gravitational lensing or by the dynamics of baryonic test particles such as stars) is expected to be modified in a redshift-dependent way. In order to study such effect we take advantage of the presence of several halos completely inside the high-resolution region, which allows us to perform a statistical analysis by stacking the profiles of all the individual halos. In order to do this  we rescaled the radial coordinates of each profiles in units of the corresponding $R_{200}$, \ie the radius enclosing a mean mass overdensity equal to $200$ times the mean density. The halos suitable to be analysed are 15, with masses between $10^{10} M_{\odot}/h$ and $10^{11} M_{\odot}/h$ at $z=4$. The errors on the stacked profile have been estimated using the variance between the individual profiles.

\begin{figure}[!t]
\centering
\includegraphics[width = \textwidth]{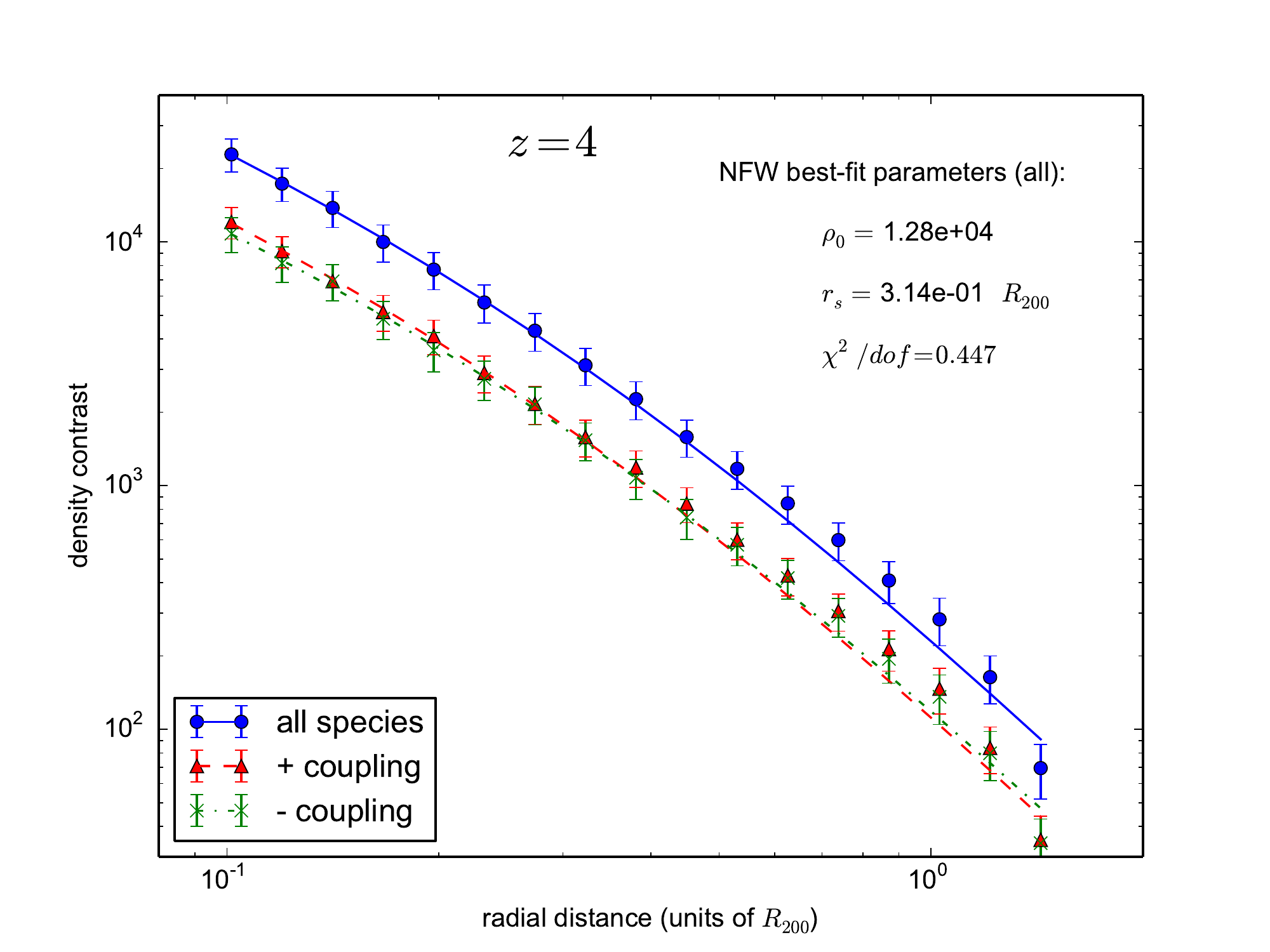}
\caption{The total stacked profile obtained for the uncoupled high-resolution simulations (blue points) at redshift $z=4$ and the best-fit to a NFW profile (blue solid line), along with the stacked profiles for the two \DM species (red triangles and green crosses) and their NFW best-fits (red dashed line and green dot-dashed line) at the same redshift. Each \DM species follows a NFW profile and thus also their sum. This is expected in the uncoupled model, since the \DM interaction is switched off, recovering the standard \LCDM cosmology.}
\label{NFW-fit}
\end{figure}

As a sanity check, we start by discussing the results obtained for the \LCDM simulation, for which no effect of the coupling is expected to occur. The goodness of our fiducial \LCDM simulations is tested by fitting the total \DM profile with a NFW profile \cite{NFW}:
\begin{equation}
\frac{\rho(r)}{\mean{\rho}_{box}} = \frac{\rho_0}{\frac{r}{r_c} \left( 1 + \frac{r}{r_c} \right)^2}\ ,
\end{equation}
where $\mean{\rho}_{box}$ is the mean density of the simulation, $\rho_0$ is a dimensionless concentration parameter equal to four times the concentration at the critical radius $r_c$, \ie where the slope of the profiles changes from the outer one $\rho \propto r^{-3}$ to the inner one $\rho \propto r^{-1}$. The best-fit is shown in Figure \ref{NFW-fit} for $z = 4$, where the stacked profile obtained in the uncoupled high-resolution simulations is shown {for the total matter density (blue dots) and for the two matter species separately (red triangles and green crosses), while the} best-fit to a NFW profile is shown using {blue solid, red dashed, and green dot-dashed lines for the same three cases, respectively}. The reduced $\chi^2$ obtained in the latter case is $\chi^2 / dof = 0.447$. A good agreement between the total matter profile derived from the \emph{uncoupled} simulations and the NFW profile is found for all the redshifts (not shown) between $z=7$ and $z=4$, with typical values of the $\chi^2$ below unity. We are therefore confident that our simulations do not introduce any artificial effect on the radial profiles, and that all the effects that we will describe below are then purely due to the McDE cosmology adopted. Moreover, we notice that each of the two \DM species separately follows a NFW density distribution. This is expected since in the uncoupled model the \DM interaction is switched off ($\beta =0$), and the two CDM particle species are then indistinguishable.

\begin{figure}[!t]
\begin{minipage}{0.5\textwidth}
\includegraphics[width=\textwidth]{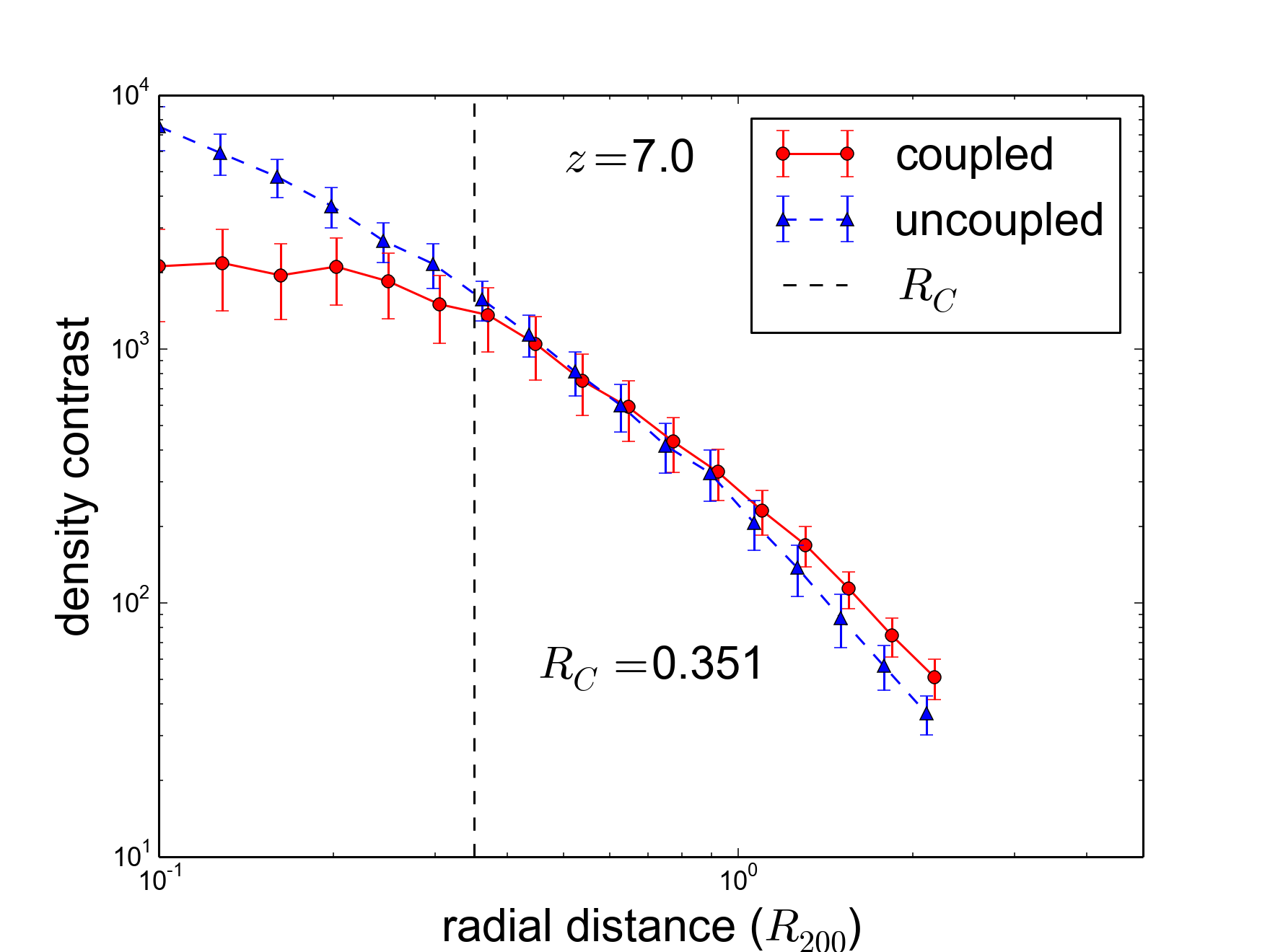}
\vspace{0.01cm}
\\
\includegraphics[width=\textwidth]{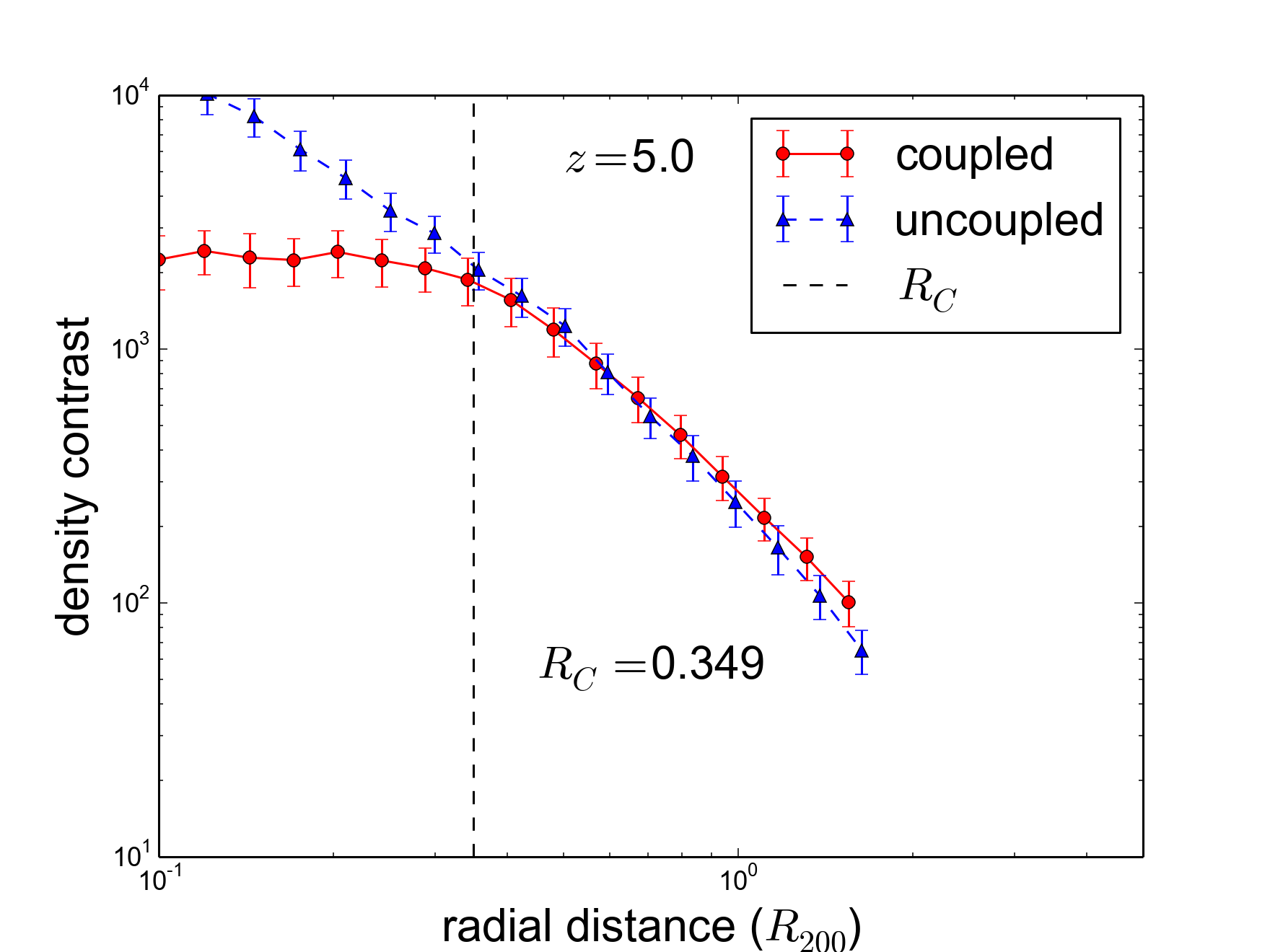}
\end{minipage}
\begin{minipage}{0.5\textwidth}
\includegraphics[width=\textwidth]{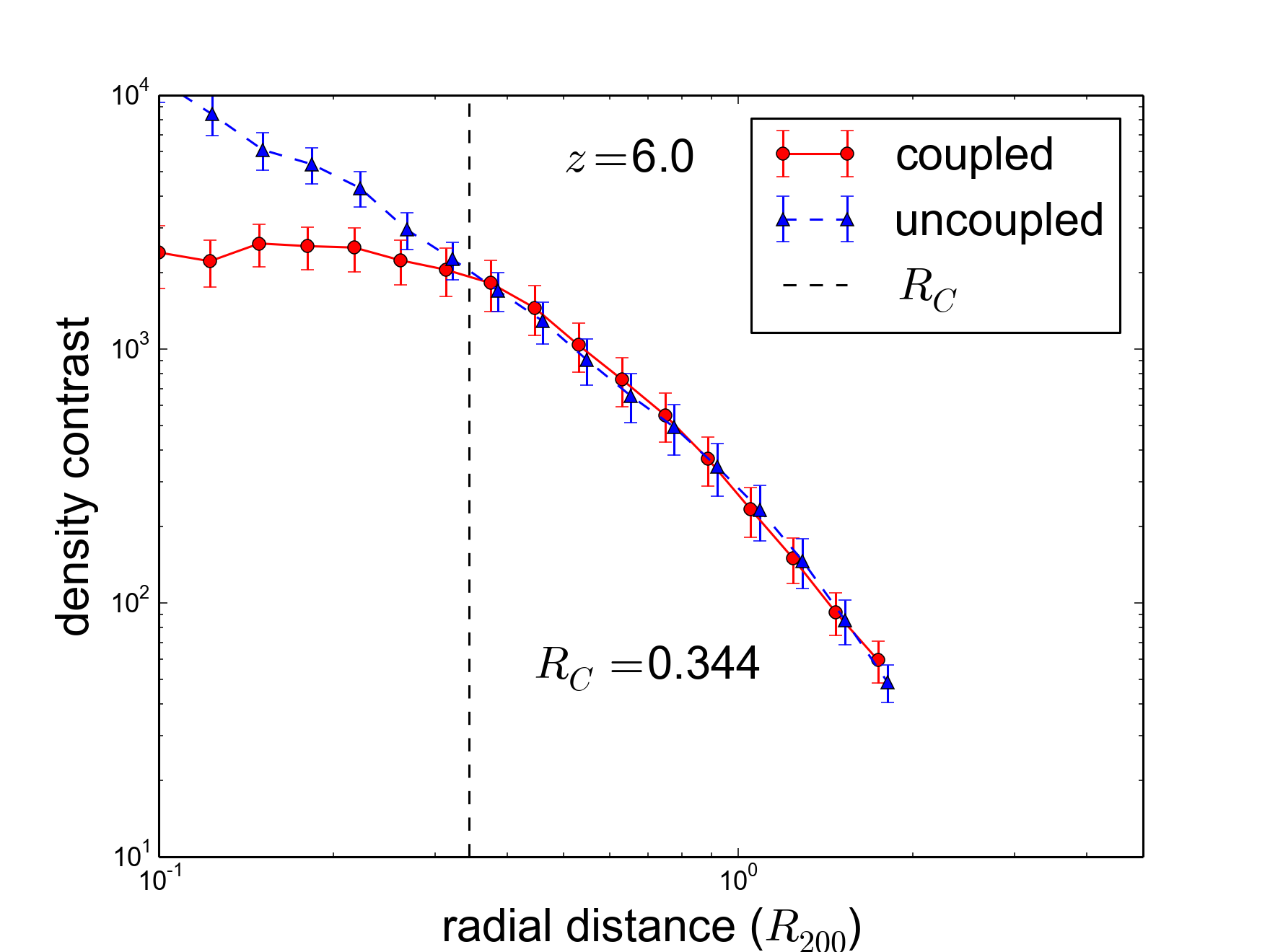}
\vspace{0.01cm}
\\
\includegraphics[width=\textwidth]{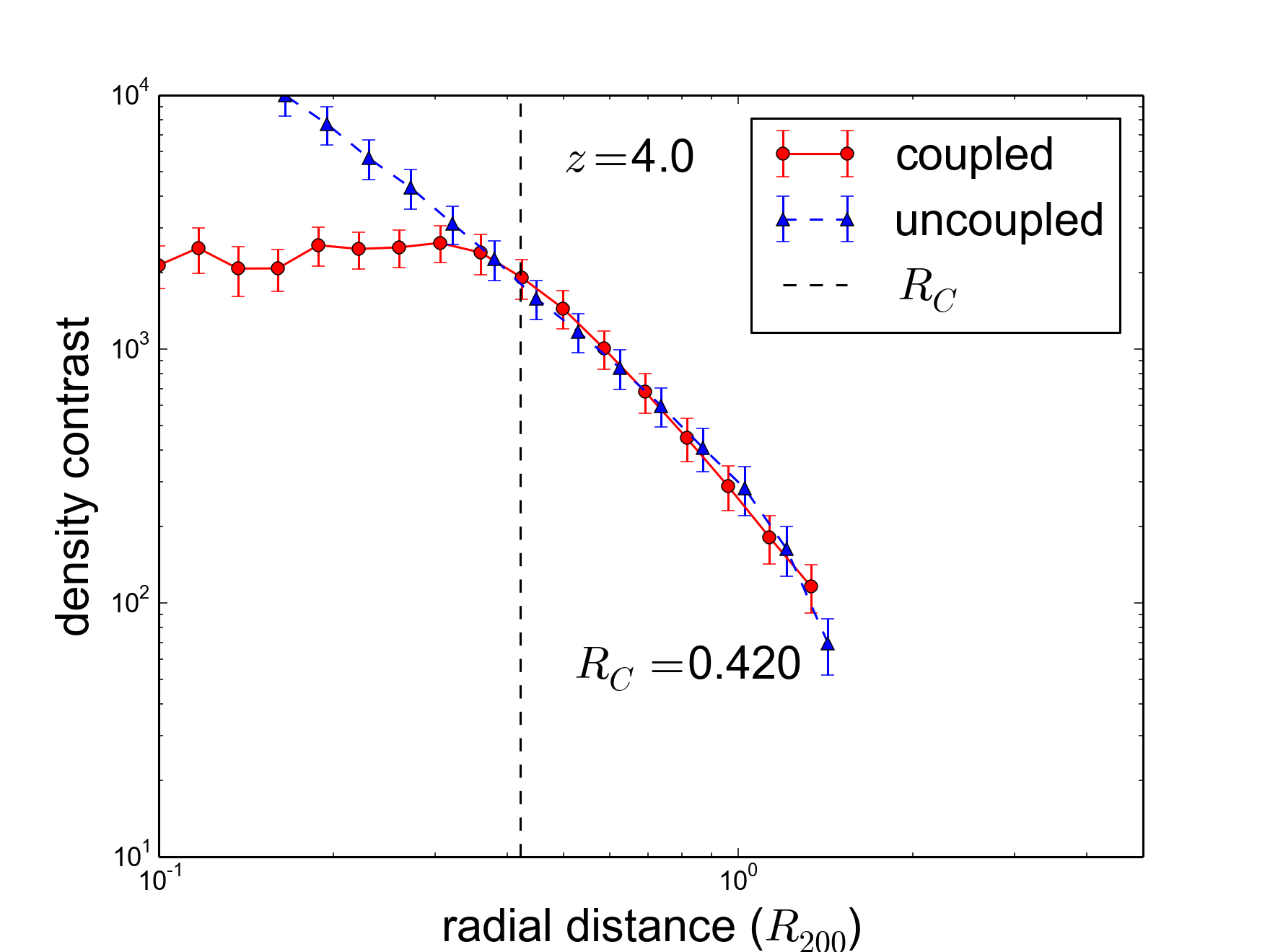}
\end{minipage}
\caption{The density constrast profiles as a function of the distance from the centre (in units of the virial radius $R_{200}$) obtained by stacking the individual profiles of the 15 halos residing within the high-resolution regions of the simulations. The different panels refer to different redshits, as labeled.}
\label{rdp_evolution}
\end{figure}

In Figure \ref{rdp_evolution} the stacked radial density profile obtained is shown at different redshifts for both the uncoupled (\LCDM, blue dashed line and triangles) and coupled (McDE, red solid line and dots) models. The first striking feature clearly visible is the appearance of a cored profile in the coupled case. In particular, at high redshift (\eg the top left panel) the radius of the core is smaller than at lower redshift, \eg at $z=4$ (bottom right panel), where the core extends to larger radii (up to $\sim 0.5R_{200}$), due to the advanced state of segregation which increases the separation between the subhalos. Consequently in the region near the centre of the halo the density is lowered with respect to the \LCDM case.

Turning to a more quantitative analysis of the profiles, we can distinguish two distinct regions: the outer part of the halo is almost indistinguishable in the two cosmologies probed with our simulations, while the inner part features striking differences. The outer part is in good agreement with a power-law 
with slope $\alpha = 3$ both for the McDE and the \LCDM simulations. On the contrary, the inner profile shows a core of roughly constant density in the coupled model whilst in the uncoupled one it closely follows the usually adopted NFW profile. 

\begin{figure}[!t]
\centering
\includegraphics[width = \textwidth]{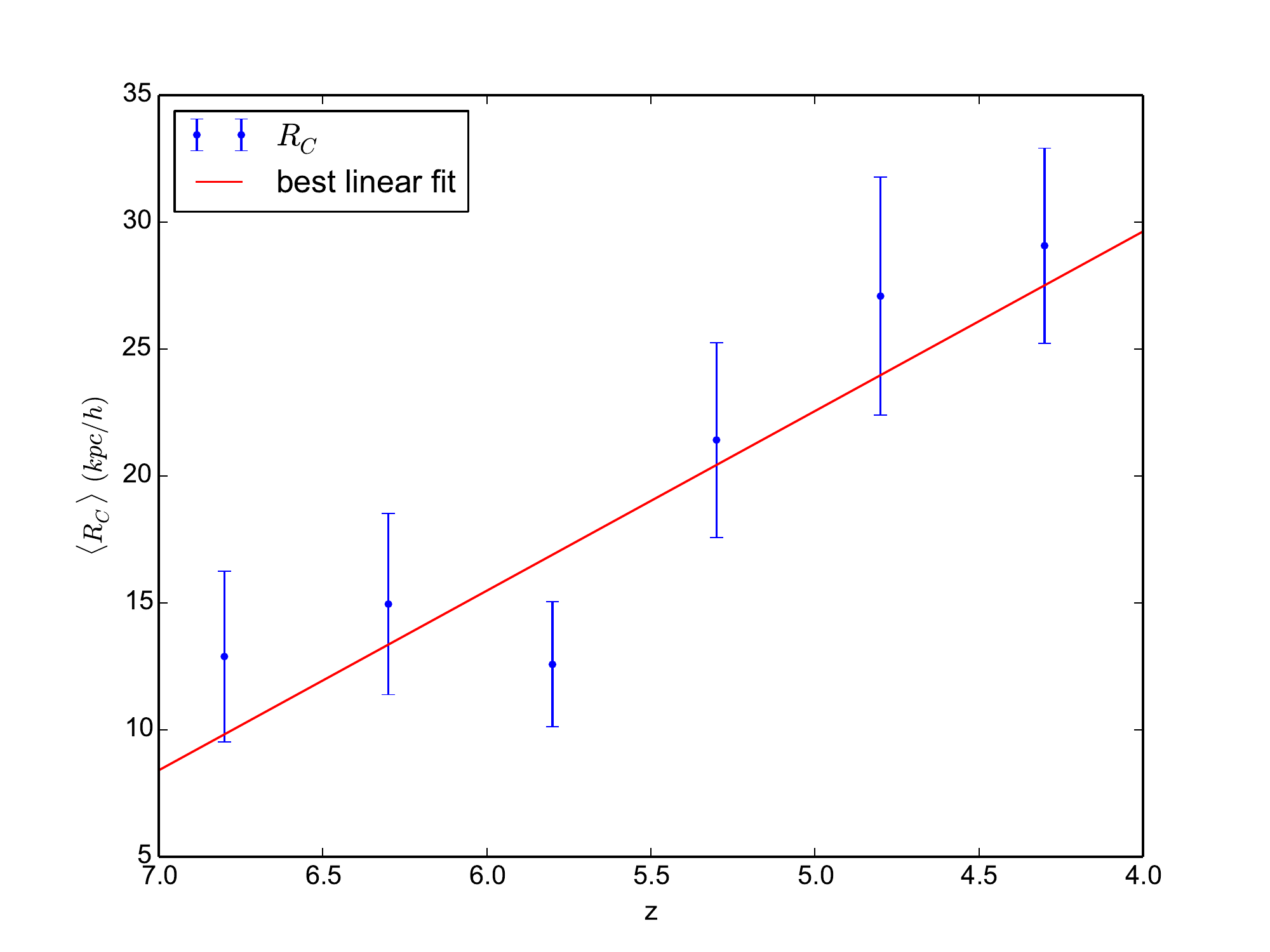}
\caption{The core radius $R_C$ as a function of redshift. The errors are estimated using the Jackknife sampling. The solid line is the best linear fit to the points.}
\label{Rc_vs_z}
\end{figure}

As briefly discussed above, in the McDE cosmology the inner part of the profile shows a slow evolution with time. In particular the core radius moves toward larger values as the redshift decreases. Such effect is qualitatively appreciable in Figure \ref{Rc_vs_z}, where the core radius $R_c$ is shown as a function of the redshift $z$. The former is estimated as the intersection between the two power-laws which best-fit the inner and outer part of the profile in the McDE model. In order to reduce the statistical noise, \emph{only} in the analysis of the core radius, the profiles were also stacked in redshift bins of width $\Delta z = 0.5$. The errors are estimated using a Jackknife sampling. In order to allow a better and easier understanding, the value obtained for the core radius in the stacked profile has been translated back to $kpc/h$. The red solid line represents the best linear fit to the points and shows an increasing trend from high to low redshift, even if a significant scatter is visible.\\

The results presented in this section allow us to draw two main conclusions on the phenomenological viability of the specific McDE model under investigation.
On the one hand, the segregation process typical of the McDE model seems to put tight constraints on the coupling strength, as the predicted bimodal density distribution is not observed in real data. Thus, the coupling strength should be sufficiently low to not produce a significant segregation at  $z=0$. On the other hand, the extended period of time over which the flattened total density profiles  seem to persist in the simulations clearly seems to suggest that a lower value of the coupling could sustain a flattened profile down to lower redshifts (and possibly until the present) before the separation of the two underlying density peaks could be resolved by observations. 

\section{Analytical Model of Halo Segregation}
\label{amodel}
In order to further test the observed segregation and its physical interpretation, we built a simple toy model to \emph{predict} the radial density profile in the McDE cosmology. For $\beta = \beta _{G}$ the segregation is expected to occur only as a consequence of the different drag terms for the two \DM species (see Eqs. \ref{gf_plus},\ref{gf_minus}), while each particle type evolves like in a standard cosmology with a gravitational constant twice as large as Newton's constant: $G_{\rm eff} = G(1+4\beta _{G}^{2}/3)=2G$. Therefore, when considering a single species of \DM, its density profile is not expected to differ from the one obtained in the \LCDM cosmology. 

Thus, in our toy model we consider a density distribution obtained by overlapping two NFW profiles (for the two CDM species) with a (variable) offset between their centres, \ie
\begin{equation}
\label{toy}
\rho(\mathbf{r}) = \rho_{NFW}(\mathbf{r}) + \rho_{NFW}(\mathbf{\Delta} - \mathbf{r})\ ,
\end{equation}
where we assumed one of the NFW profiles is centred in the origin and the other one in $\mathbf{\Delta}$. Then, we compute the total profile assuming as center of the halo the middle point between the centres  of the individual NFW profiles. The analytical model is completely independent of the McDE profile, \ie it was by no mean calibrated against it. The only inputs for our model come from the uncoupled total-matter density profile best-fit parameters (see Figure \ref{NFW-fit}), which were used as parameters for the two NFW profiles in the density distribution (after halving $\rho_0$).

\begin{figure}[!t]
\centering
\includegraphics[width = \textwidth]{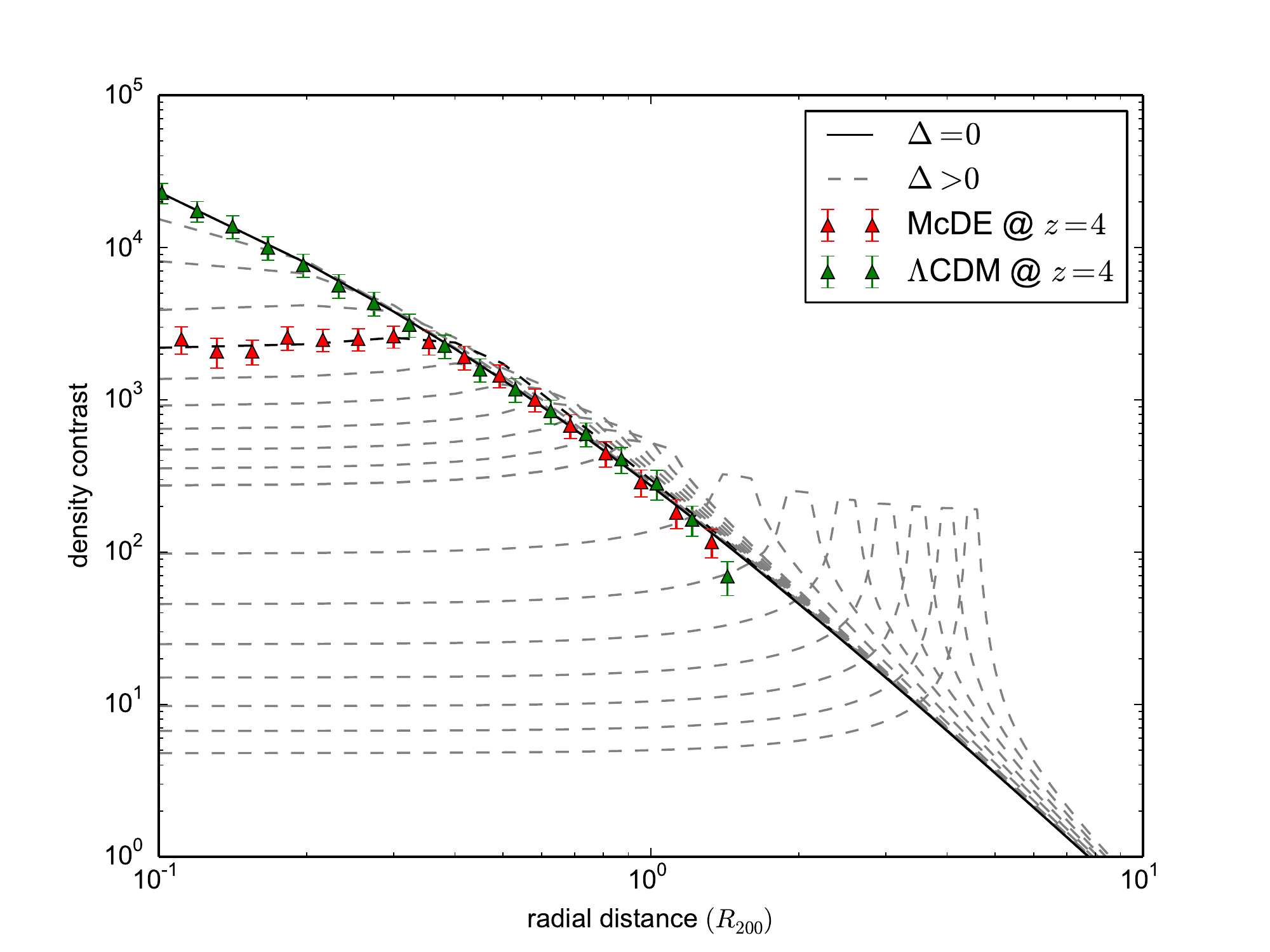}
\caption{The predicted density profile based on our toy model \ref{toy} as seen from the middle point between the centres of two NFW density distributions. The solid line is the profile obtained for no offset while the dotted lines are the ones obtained for values of the offset increasing from the top line to the bottom one. The values of the offset plotted are: $0.0$, $0.1$, \dots, $1.0$, $1.5$, \dots, $5.0$. The green and red triangles represent the stacked uncoupled and coupled profiles, respectively, as extracted from our zoom-in simulations at $z=4$.}
\label{modelev}
\end{figure}

The expected profile was computed for different values of the offset $\Delta = \abs{\mathbf{\Delta}}$. The predicted profiles for different offsets are shown in Figure \ref{modelev} as dashed lines. The solid line is the profile predicted for $\Delta = 0$, \ie the one expected in the \LCDM cosmology. The green and red dots represent the stacked profiles obtained from the simulations in the uncoupled and coupled case, respectively.

As we can see in the plot, the uncoupled profile matches the one predicted for $\Delta = 0$, but this is expected since the toy model was calibrated against it. Remarkably, also the coupled profile matches quite well the profile predicted for a superposition of two offset NFW shapes. Such a remarkable match strongly back up both the theoretical interpretation of the segregation process and the results of the simulations described in this work. Moreover, it is possible to use the analytical model to estimate the offset which best fits the stacked profile from our simulations and compare it with the average value for the profiles stacked. The mean offset computed from the simulations is $\mean{\Delta}_{sim} = 0.837 \, R_{200}$, to be compared with the one computed from the toy model minimizing the $\chi^2$ (highlighted in Figure \ref{modelev} with a thick dashed line), which is $\mean{\Delta}_{model} = 0.792 \, R_{200}$. Given the simplicity of the analytical model, which overlooks \eg the redshift evolution of the \DM particles' mass, the two values match remarkably well.

\section{Discussion and Conclusions}
\label{concl}

We have presented the outcomes of the first high-resolution zoom-in simulations ever performed for the Multi-coupled Dark Energy scenario. The latter is a cosmological model characterised by two distinct species of Cold Dark Matter particles that interact with opposite couplings to a light scalar field playing the role of the Dark Energy. The McDE model represents an appealing alternative to the standard $\Lambda $CDM cosmology as it features a highly non-trivial phenomenology at the level of linear and (mainly) non-linear structure formation processes while -- thanks to a dynamical screening during matter domination -- it is almost completely indistinguishable from the standard model at the level of background geometrical observables.

The most striking signature of the model, that was first highlighted {in B13 (and subsequently further investigated in \citep[][]{Baldi_2014})} by means of cosmological intermediate-resolution simulations, is the so-called {\em halo segregation} process, i.e. the tendency of collapsed objects to split (at low redshifts) into pairs of ``mirror" structures dominated by only one species of CDM particles. Such phenomenon arises as a consequence of the effective violation of the Weak Equivalence Principle occurring in Multi-coupled Dark Energy cosmologies. Previous numerical simulations have provided a statistical and dynamical characterisation of the {\em halo segregation} process, but lacked the spatial resolution to resolve the early stages of the segregation and to investigate the consequences of this phenomenon on the observable properties (such as the total density profile) of individual halos.

In the present work, we have made the further step of simulating at sufficiently high resolution a set of individual halos during the whole phase of segregation. This has been accomplished by means of the zoom-in re-simulation technique that allows to increase the numerical resolution around a specific target halo without discarding the effects of the large-scale gravitational field that is self-consistently evolved at much lower resolution. To this end, we have made use of a new code -- the \zinco code, which we make publicly available\footnote{\url{https://github.com/EGaraldi/ZInCo}} -- for the setup of zoomed-in initial conditions starting from existing high-resolution initial conditions files. The details of the \zinco code are described in the Appendix \ref{ZInCo}.

By means of a suitably modified version of the {\small GADGET} code, we have evolved these zoomed initial conditions from $z=99$ to $z=4$ both in the context of the standard $\Lambda $CDM cosmology and for a Multi-coupled Dark Energy model characterised by a coupling strength corresponding to scalar fifth-forces with the same strength as standard gravity. Therefore, in our simulations CDM particles of opposite types do not interact with each other in any way, since the repulsive fifth-force exactly cancels their standard gravitational attraction. Although we ran the zoomed simulations only for two different target halos, as a consequence of our conservative choice of the high-resolution regions in the initial conditions we could actually rely on other 13 halos fully residing in the high-resolution region, so that we could perform a stacking of their density profiles and compare the stacked profiles in the two models.

\noindent The main results of our analysis are the following:
\begin{itemize}
\item[I:] The segregation process starts at much higher redshifts than expected based on the results of previous simulations. This is due to the lower resolution of these previous studies that did not allow to resolve the separation of the two distinct density peaks that indicate the onset of the segregation process. Our results show that the two CDM components of a collapsed halo start to segregate already at redshifts as high as $z\sim 7$;
\item[II:] As a consequence of the segregation, and of the progressive separation of the two individual structures, the total matter density profile evolves from the cuspy NFW shape (that we recover, as expected, for the $\Lambda $CDM halos, and that should represent the initial state of a collapsed object also in the coupled model) to a cored profile;
\item[III:] The size of the core increases for decreasing redshift, as a consequence of the progressive separation of the two ``mirror" structures arising from the segregation;
\item[IV:] The shape of the cumulative density profile can be accurately predicted by a simple toy model featuring a superposition of two individual NFW profiles with the same overdensity parameters shifted from one another by a redshift-dependent offset.
\end{itemize}

Our simulations therefore provide a clear insight to the process of halo segregation for the simplified case of an exactly gravitational coupling in McDE cosmologies. This special condition appears to be clearly ruled out since mirror structures are not observed in real data. However, our simplified analysis provides a clear understanding of the mechanisms leading to the halo segregation and to the formation of a cored profile in the total matter density starting from an initially cuspy NFW profile. 

Therefore, our results pave the way for the investigation of the halo segregation process at progressively smaller values of the coupling, which are expected to provide a more realistic evolution of cosmic structures and possibly determine the formation of cored halos at lower redshift (desirably even down to $z=0$) without ever evolving into directly observable mirror objects.

\section*{Acknowledgments}
We acknowledge the CINECA award under the ISCRA initiative, for the availability of high performance computing resources and support. EG acknowledges support from the Collaborative Research Centre (SFB) 956, sub-project C4, funded by the Deutsche Forschungsgemeinschaft (DFG). MB acknowledges support from the Italian Ministry for Education, University and Research
(MIUR) through the SIR individual grant SIMCODE, project number RBSI14P4IH.
LM acknowledges the grants ASI n. I/023/12/0 ``Attivit\`a relative alla fase B2/C per la missione Euclid'', MIUR PRIN 2010-11 ``The dark Universe and the cosmic evolution of baryons'' and PRIN INAF 2012 ``The Universe in the box: multiscale simulations of cosmic structure''. 

\appendix
\section{The ZInCo code}
\label{ZInCo}
In order to produce initial conditions (ICs) suitable for the simulations described in this Paper, we developed a code called \zinco (a simple acronym for \emph{\textbf{Z}oomed \textbf{In}itial \textbf{Co}nditions}). Here we summarise its key features. The code is able to handle ICs for the cosmological N-body code \gadget (and newer versions), last described in \citep{gadget} and \cite{gadget-2}.

The main purpose of the code is to \emph{manipulate} existing initial conditions in order to allow different flavours of zoom-in simulations, rather then producing new ICs from scratch. This approach offers two main advantages: firstly, it can take advantage of the wide number of codes already existent which produce ICs featuring a broad range of different cosmologies, since each set of ICs compatible with the cosmological code \gadget can be processed. Secondly, for the same reason, it can be used also on existing ICs even in the unlikely case nothing is known about their properties. In this sense, it is somewhat \virgolette{backward compatible}. The code is able to manipulate initial conditions with multiple types of particles, unlike the vast majority of zoom-in ICs codes available, preserving their properties and random field.

The approach chosen for the code is to start from a high-resolution set of initial conditions, corresponding to the highest resolution desired, and then reduce their resolution (\emph{dilute}) outside the region(s) of interest. The original high-resolution ICs are therefore never actually used any N-body integration. Such procedure allows a full customization of the ICs generator, which is completely independent of \zinco. Moreover, it ensures that the highest resolution particles (\ie the ones of greatest interest) are exactly the same as the original ICs, since they are copied unmodified to the final ICs.

The code builds up on two main elements: \emph{resolution regions}, which are theoretical spherical regions defining a sub-volume of the simulation box and its corresponding resolution (in the output ICs); and \emph{cubic domains}, each of them defined by its side and belonging to one and only one resolution region, corresponding to a single low-resolution output particle representing all the particles enclosed in the domain. These two elements represent the input and the output of the code, respectively. Since \gadget allows for 6 different types of particles, each one of them is processed separately by \zinco.

\begin{figure}[!t]
\centering
\includegraphics[height = 0.4\textwidth]{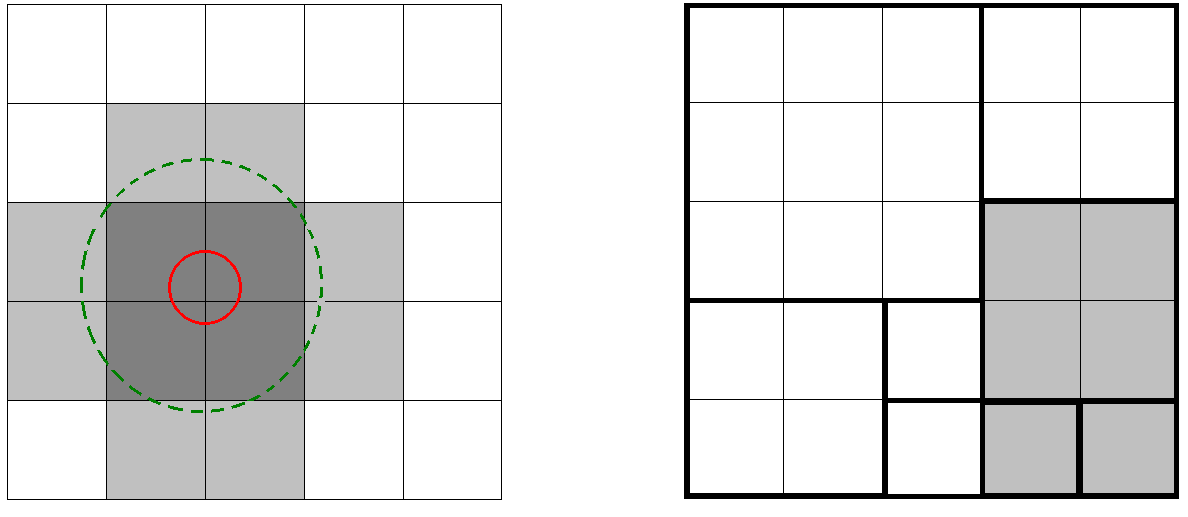}
\caption{Schematic representation of the main steps of the \zinco algorithm, shown in 2 dimensions for simplicity. The whole simulation box is divided in $4^2$ cubic cells in this example. \emph{Left}: Assignment of the cells to the highest-level resolution regions they overlap with. In the example above there are 3 regions: the red solid and green dashed circles represent the highest- and mid-level regions, respectively, while the lowest resolution region correspond to the whole box. Dark grey shaded, light grey shaded and white cells belong to the highest-, the mid- and the lowest-resolution level, respectively. \emph{Right}: Example of cubic domain assignment for just 2 resolution levels (light gray shaded and white cells for high and low level, respectively). The code was instructed to merge $2^2$ cells for the high level and $3^2$ for the low level. Thin lines contour the cells, while thicker lines contour the computed domains. Note how smaller (but always cubic) domains can occur when there are not enough unprocessed cells to merge in a region (\eg white $2^2$ and $1^2$ and light grey $1^2$ domains).}
\label{slicing}
\end{figure}

The algorithm at the core of the code is conceptually straightforward. Firstly the whole simulation box is divided in cubic cells. The user must provide a number of resolution regions. The resolution is assigned using an integer parameter, $m$, which is the side of the cubic domains (belonging to a given region) in units of cubic cells, \ie a low-resolution particle is produced merging $m^3$ neighbouring cells. The code loops through all the cubic cells assigning them to the highest-level\footnote{Note here that highest-level usually means highest-resolution, but this is not enforced by the code.} resolution region they overlap with (left panel of Figure \ref{slicing}). Then, \zinco creates cubic domains grouping together cells which belong to the same level. When there are not enough neighbour cells, the $m$ parameter is decreased by one and the grouping repeated for the unprocessed cells, until all the cells are assigned to a domain (right panel of Figure \ref{slicing}). The resulting iso-resolution surfaces will appear as spheres made up of cubic pieces. However they are ensured to enclose the whole corresponding resolution region defined by the user. Each cubic domain is now processed separately replacing the particles inside it with their centre of mass. The code can be instructed to not merge the particles inside a given resolution region. Instead the particles inside the cubic domains of the given region can be discarded, producing an empty region, or copied unprocessed, \ie with their original (high) resolution, to the new ICs.

\begin{figure}[tp]
\begin{center}
\includegraphics[height=0.27\textheight]{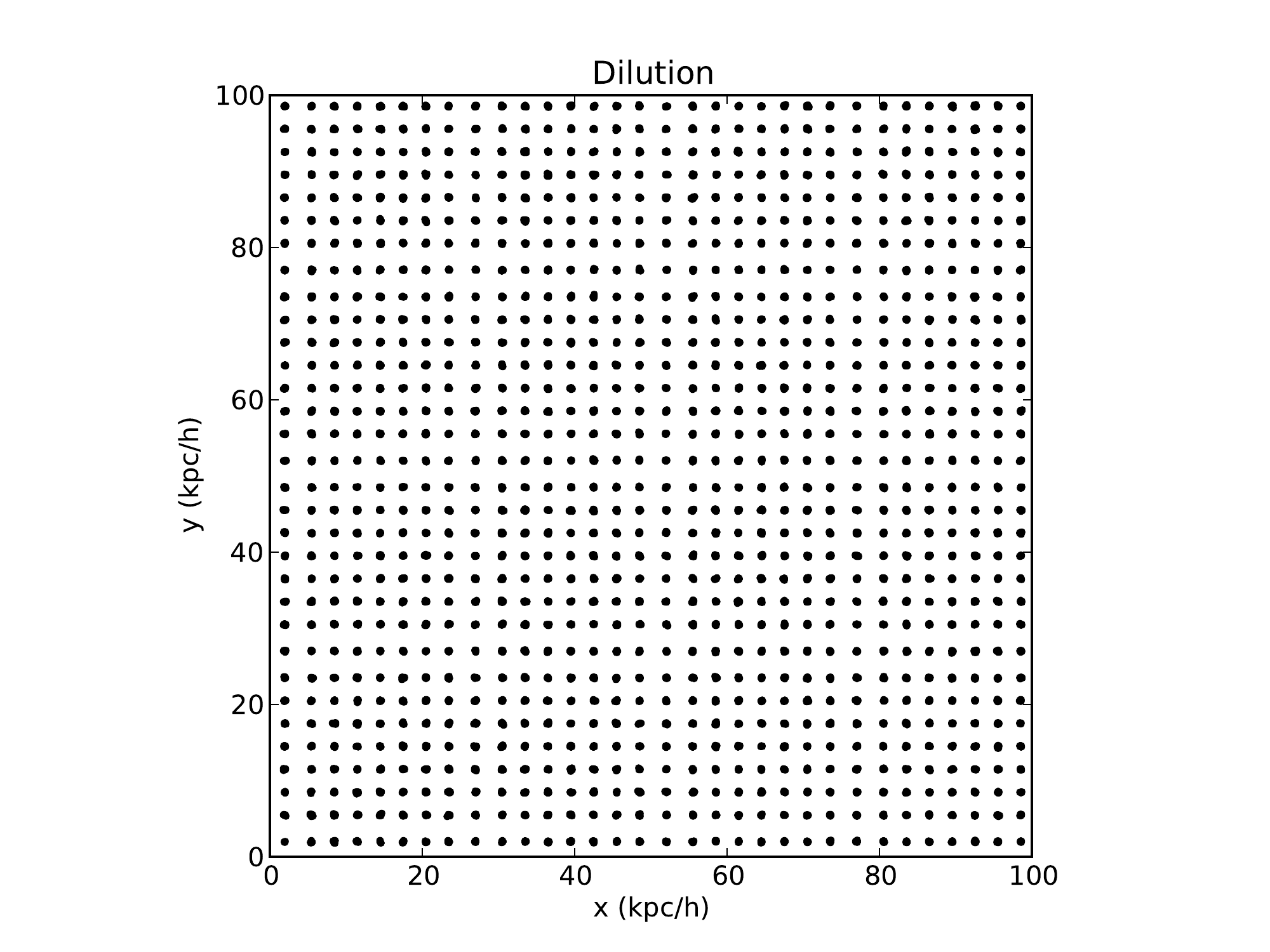}
\vspace{0.01cm}
\\
\includegraphics[height=0.27\textheight]{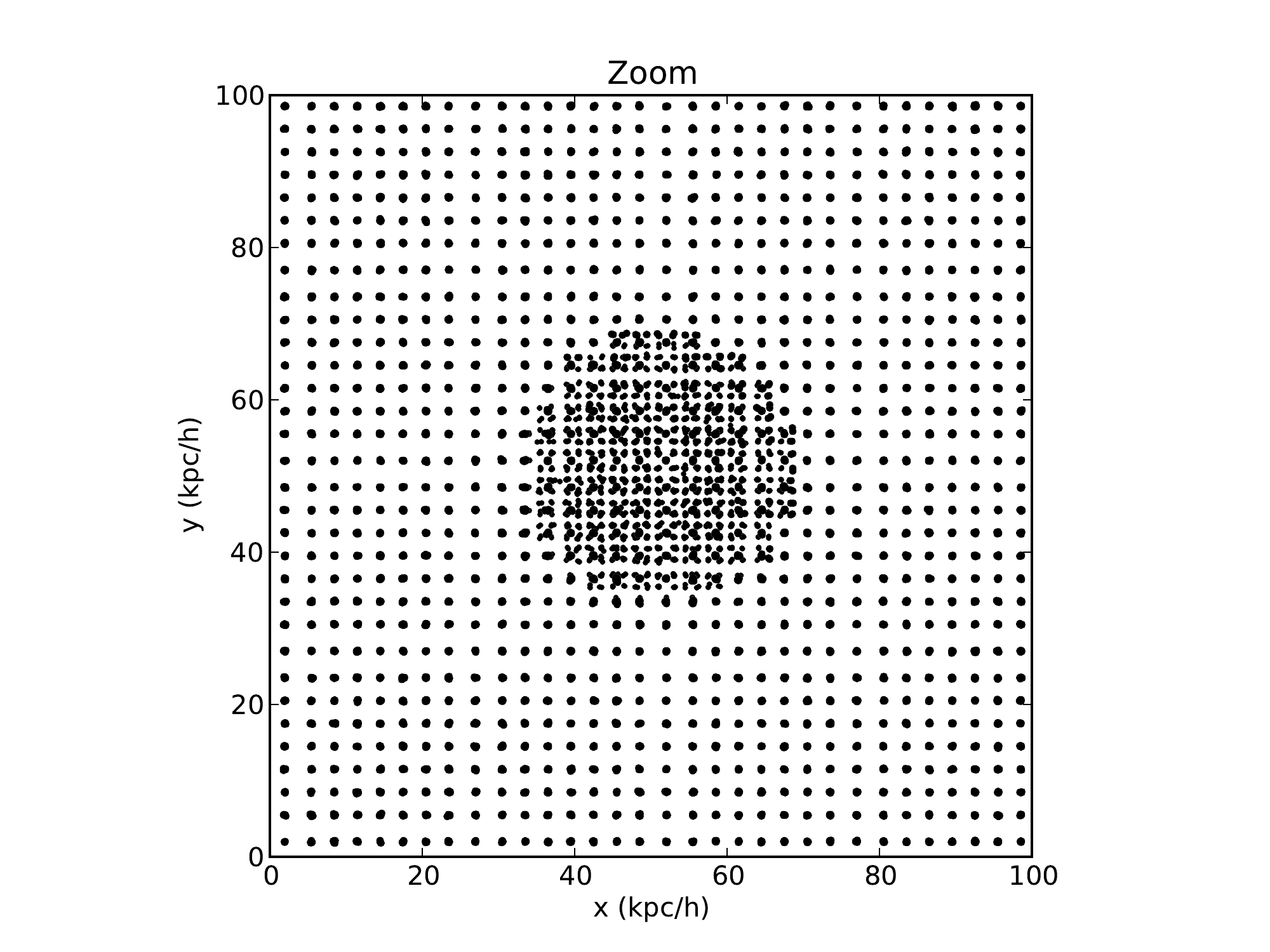}
\vspace{0.01cm}
\\
\includegraphics[height=0.27\textheight]{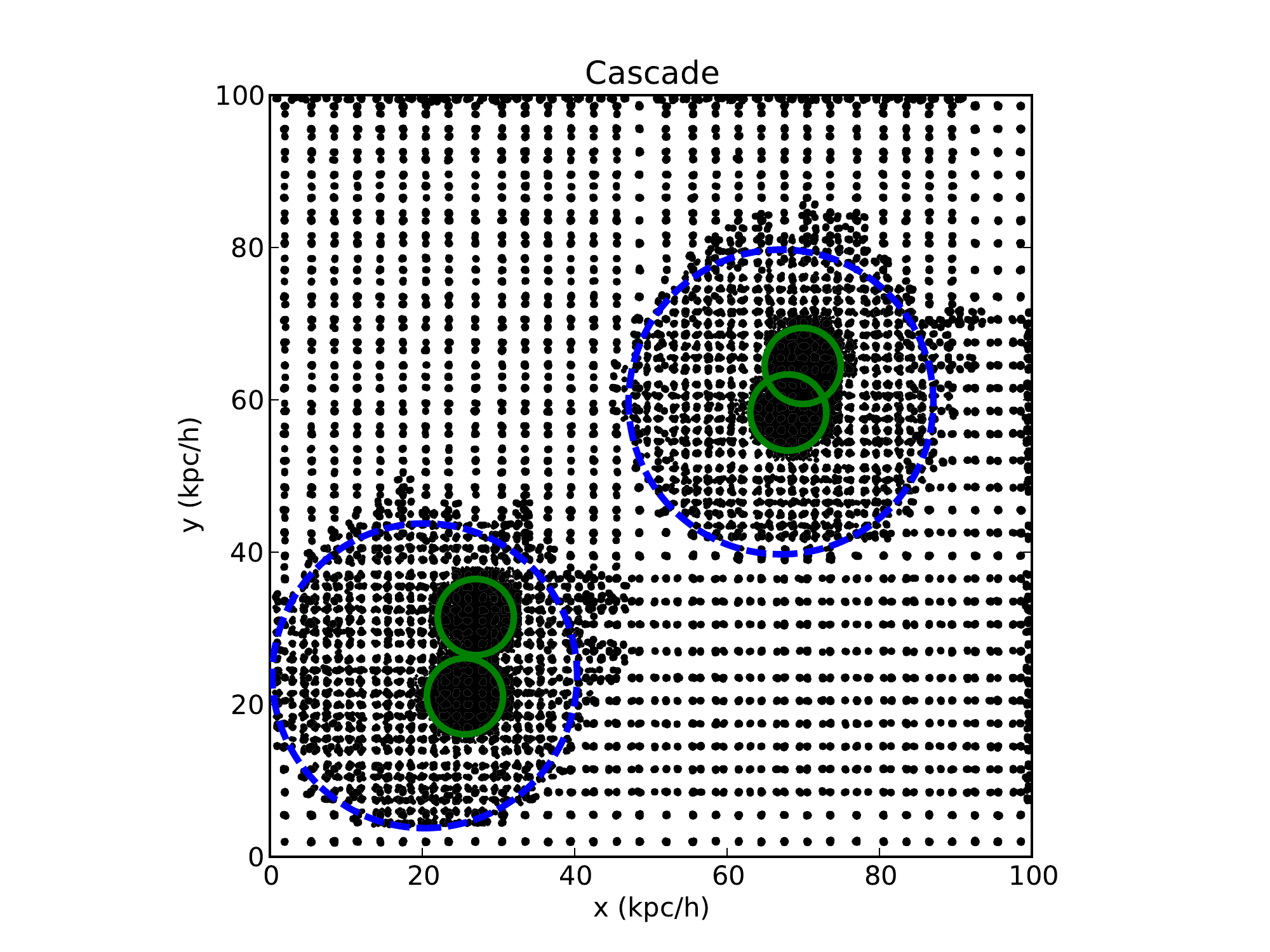}
\end{center}
\caption{Three different manipulations of the same set of initial conditions performed by \zinco. Each panel represents a slice of thickness $25 Mpc/h$ from a box of side $100 Mpc/h$. \emph{Top}: uniform dilution of the ICs. \emph{Middle}: Zoom on a single region of interest (in this example located at the centre of the box). Different dot sizes represent different resolution levels (the bigger the dot, the lower the resolution). \emph{Bottom}: Cascade ICs made of resolution regions randomly located in the box (in this example 3 different resolution levels). The blue and green circles are the medium- and high-resolution bubbles, respectively.}
\label{zinco}
\end{figure}

\zinco allows for three different manipulations of the ICs, namely:
\begin{itemize}
\item \emph{dilution}, where the resolution is uniformly reduced, \ie only a single resolution region exists (useful for preparatory runs, top panel of Figure \ref{zinco});
\item \emph{zoom}, which is used to zoom on a single region of interest. The number of resolution regions is chosen by the user, but they must share the same center; their centre and radii can be specified in a snapshot of the simulation instead of in the original ICs: in this case, the code tracks the particles in the regions back to the ICs and update the resolution regions consequently (middle panel of Figure \ref{zinco});
\item \emph{cascade}, which is used to produce recursively smaller and more-resolved \virgolette{bubbles}; the number of regions can be specified by the user for each level of resolution, while their positions are set randomly (bottom panel of Figure \ref{zinco}).
\end{itemize}

Among the most important parameters that can be set, there are: the number of cubic cells the simulation box is divided in, which ultimately controls the resolution in the new ICs produced; number, center (only \emph{zoom} run) and radius of the resolution regions; the number of different particle species available in the original ICs; how many bubbles each level will contain (\emph{cascade} run only); the resolution of each level by mean of the parameter $m$.

\begin{figure}[!pt]
\centering
\includegraphics[width=0.8\textwidth]{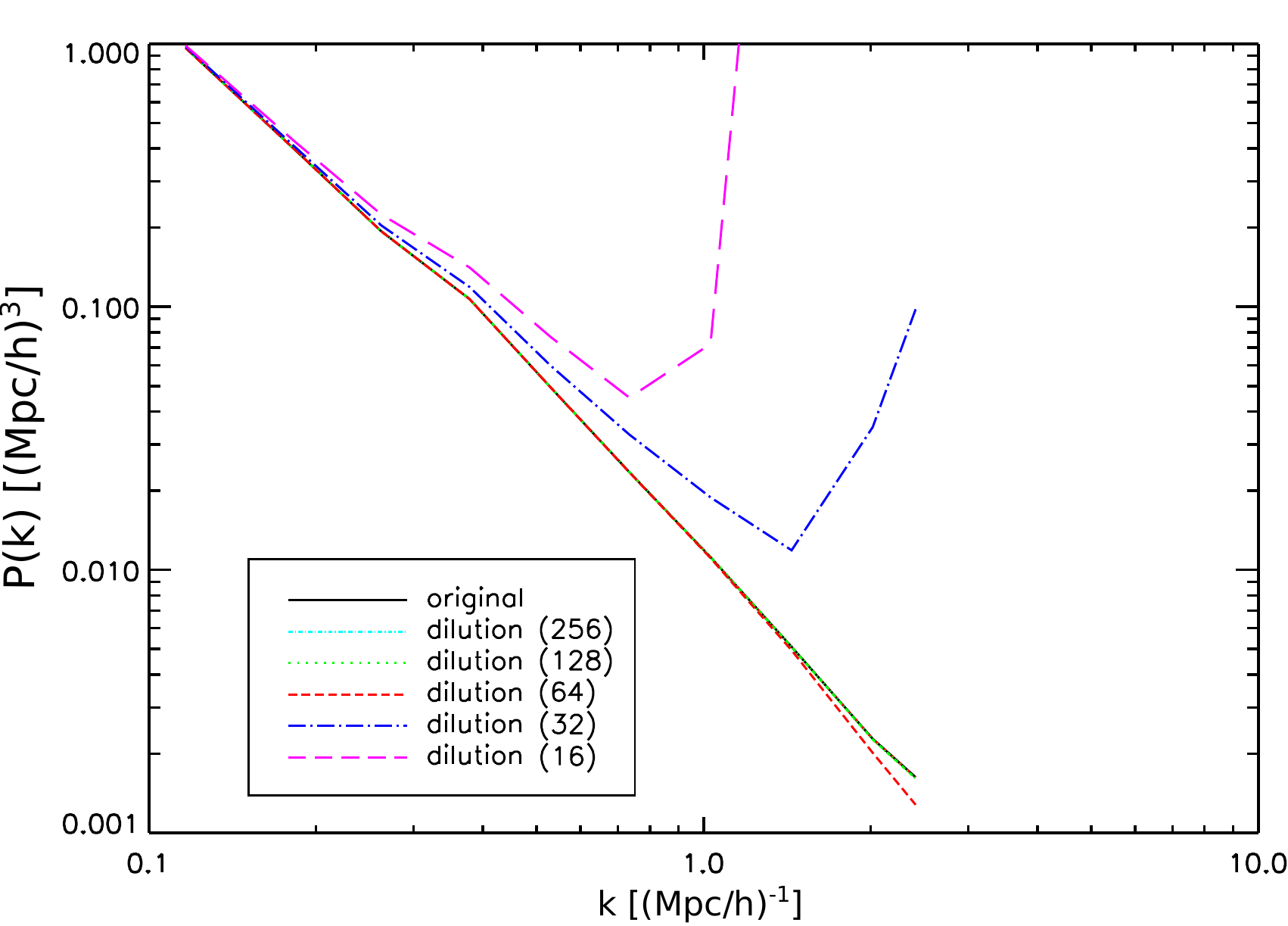}
\caption{Comparison between the computed power spectrum $P(k)$ of the ICs field for different levels of dilution, as a function of the wavenumber $k$. The  original initial conditions features $512^3$ particles, while the numbers in the legend are the number of particles in each dimension for the diluted ICs. The power spectrum is preserved in the dilution procedure up to wavenumbers where the shot noise plays a role.}
\label{zinco_test}
\end{figure}

As a simple example of the tests we performed for the code, in Figure \ref{zinco_test} we show the power spectrum computed for different dilutions. Each one of them is obtained from the same set of initial conditions (with $512^3$ particles). As expected, the original power spectrum is preserved in the dilution procedure and therefore differs only at small scales, where the shot noise $\Delta_{SHOT} \propto \frac{1}{N} k^3$ (with $N$ as the number of particles), becomes relevant.

\bibliographystyle{JHEP}
\bibliography{baldi_bibliography}

\label{lastpage}

\end{document}